\documentclass[a4paper,11pt]{article}
\pdfoutput=1 

\usepackage{jheppub} 

\usepackage[T1]{fontenc} 
\usepackage{amsmath}
\usepackage{graphicx}
\usepackage{booktabs}
\usepackage{multirow}
\usepackage{xspace}
\usepackage{commands}
\usepackage{subfigure,graphicx}

\usepackage[colorinlistoftodos]{todonotes}

\title{\boldmath Measuring the Higgs self-coupling via Higgs-pair production at a 100 TeV p-p collider}

\author[a]{Michelangelo~L.~Mangano}
\author[b]{\!\!,Giacomo~Ortona}
\author[a]{and Michele~Selvaggi}

\affiliation[a]{CERN, CH-1211 Geneva 23, Switzerland}
\affiliation[b]{INFN Sezione di Torino, via P. Giuria 1, 10125, Torino, Italy}

\emailAdd{michelangelo.mangano@cern.ch}
\emailAdd{ortona@to.infn.it}
\emailAdd{michele.selvaggi@cern.ch}

\abstract{Higgs pair production provides a unique handle for measuring the strength of the Higgs self interaction and constraining the shape of the Higgs potential. Among the proposed future facilities, a circular 100 TeV proton-proton collider  would provide the most precise measurement of this crucial quantity. 
In this work, we perform a detailed analysis of the most promising decay channels and derive the expected sensitivity of their combination, assuming an integrated luminosity of 30 ab$^{-1}$. Depending on the assumed detector performance and systematic uncertainties, we observe that the Higgs self-coupling will be measured with a precision in the range 3.4 - 7.8\% at 68\% confidence level.}

\begin{document}
\maketitle
\flushbottom

\clearpage

\section{Introduction}
\label{sec:intro}

The steady progress of the LHC experiments keeps improving our knowledge of the Higgs properties~\cite{Aad:2019mbh,Sirunyan:2018koj}. The long-term prospects for the high-luminosity phase of the LHC (HL-LHC) set important precision goals~\cite{Cepeda:2019klc}, reaching the level of few percent for several of the Higgs couplings to gauge bosons and fermions. Beyond this, the per-mille level frontier is opened by a future generation of Higgs factories~\cite{deBlas:2019rxi}. The measurement of the Higgs self-coupling, the key parameter  controlling the shape of the Higgs potential, will remain however elusive for a long time. Aside from providing clues to the deep origin of electroweak (EW) symmetry breaking (EWSB), the determination of the Higgs potential has implications for a multitude of fundamental phenomena, ranging from the nature of the EW phase transition (EWPT) in the early universe~\cite{Kajantie:1996qd}, to the (meta) stability of the EW vacuum~\cite{Cabibbo:1979ay,Hung:1979dn,Lindner:1985uk,Sher:1988mj,Degrassi:2012ry}. This measurement sets therefore a primary target among the promised guaranteed deliverables of any future collider programme. Comparative assessments of the potential of different collider options, relying on studies carried out through the years in preparation for their design studies, have recently appeared in two reports~\cite{deBlas:2019rxi,DiMicco:2019ngk}. The $\pm 50\%$ precision projected for the HL-LHC~\cite{Cepeda:2019klc} can be improved by a factor up to 2 at future $e^+e^-$  colliders~\cite{deBlas:2019rxi,Blondel:2018aan}, exploiting the impact of radiative corrections induced by the Higgs self-coupling on single-H production at several energies below the onset of on-shell Higgs-pair (HH) production~\cite{McCullough:2013rea}. The direct measurement of HH production at $\sqrt{s}\ge 1$~TeV will provide stronger, and independent, measurements, reaching 10\% and 9\% for the ILC at $\sqrt{s}=1$~TeV~\cite{Fujii:2019zll} and CLIC at $\sqrt{s}=3$~TeV~\cite{Charles:2018vfv}, respectively.  These measurements will require a longer time scale, as they will be possible only at the last stage of the proposed ILC and CLIC programmes. On these timescales, comparable or even better precision could be possible via the study of HH production at a future high-energy proton-proton (pp) collider, like the 100~TeV Future Circular Collider~\footnote{For the sake of simplicity, we shall just refer in the following to FCC-hh.} (FCC-hh~\cite{Abada:2019lih} or the SPPC~\cite{CEPCStudyGroup:2018ghi}). 

HH production in hadronic collisions has long been considered as an ideal probe of the Higgs self-coupling~\cite{Baur:2002rb,Blondel:2002nta,Gianotti:2002xx}, and much work along these lines has been done since the Higgs discovery. Some of the most recent work, in the context of future colliders, is documented in Refs.~\cite{Yao:2013ika,Barr:2014sga,Liu:2014rva,He:2015spf,Kumar:2015kca,Fuks:2015hna,Papaefstathiou:2015iba,Cao:2016zob,Bishara:2016kjn,Contino:2016spe,Banerjee:2018yxy,Goncalves:2018yva,Homiller:2018dgu,Chang:2018uwu,Biekotter:2018jzu,Borowka:2018pxx,L.Borgonovi:2642471,Li:2019uyy,Agrawal:2019bpm,Banerjee:2019jys,Park:2020yps}. The best estimates, obtained in these studies,  of the sensitivity to the Higgs self-coupling at the FCC-hh  have used the \bbaa\ decay channel, leading to an achievable precision between 5-10\%, using this channel alone. A study focusing on the \bbtata\ and \bbbb\ final states~\cite{Banerjee:2018yxy} in the boosted regime achieved a sensitivity of 8\% and 20\%, respectively. The most up-to-date result, performed by the FCC-hh collaboration~\cite{Abada:2019lih,L.Borgonovi:2642471} quotes a precision of 5-7\%, driven by the \bbaa\ channel.

The goal of the present study is to extend the scope of previous projections summarized in Ref.~\cite{Abada:2019lih} and to provide a refined and comprehensive reference for the combined prospect for the Higgs self-coupling measurement at the FCC-hh. We improve on previous studies and show that further optimization of the most sensitive Higgs decay channels using multi-variate techniques is possible. When interpreted in the framework of the Standard Model (SM), the combination of these measurements of HH production allows to reach a precision on the tri-linear Higgs self-coupling in the range $\dkl=3.4-7.8\%$, significantly improving previous estimates.

This article is organized as follows. We introduce the theoretical framework, discussing the relation between Higgs self-coupling and HH production, in Section~\ref{sec:theory}, and we present in Section~\ref{sec:simulation} the event generation tools used for this study. The detector modeling, event simulation and analysis frameworks are discussed in Section~\ref{sec:expsetup}. 
In Section~\ref{sec:strategy} we introduce the general measurement strategy and the procedure that we use for the signal extraction and to derive the expected precision on the self-coupling. The analyses of the three most sensitive decay channels \bbaa, \bbtata\ and \bbbb\ final states and their combination are presented in Section~\ref{sec:analysis}. Section~\ref{sec:conc} summarizes our results and our conclusions.

\section{The theoretical framework}
\label{sec:theory}
Perturbing the Higgs potential around its minimum, leads to the general expression:
\begin{equation}
    \mathcal{L}_{h} = \frac{1}{2}\mH^2 H^2 + \lthree H^3 + \lfour H^4,
\label{eq:potential}
\end{equation}
where $\mH$ is Higgs boson mass and $\lambda_3$ and $\lambda_4$ are respectively the trilinear and quartic Higgs self-couplings. In the SM the self-couplings are predicted to be $\lambda_3^{SM}={\mH^2}/{2v}$, $\lambda_4^{SM} = {\mH^2}/{8v^2}$, where $v$ is the vacuum expectation value (vev) of the Higgs field. The Higgs vev is known from its relation to Fermi constant, $v=(\sqrt{2}G_F)^{-1/2}=246$~GeV, and the discovery of the Higgs particle at the LHC~\cite{Aad:2012tfa,Chatrchyan:2012ufa} has fixed the last remaining free parameter of the SM, the Higgs mass \mH~\cite{Aad:2015zhl}. Beyond the SM, corrections to $\lthree$ and $\lfour$, as well as higher-order terms, are possible. 

To this day, large departures from the SM potential are perfectly compatible with current observations~\cite{Aad:2019uzh,Sirunyan:2018two}. This makes it possible, for example, to contemplate BSM models where the modified Higgs potential allows for a strong first order EW phase transition (SFOPT) in the early universe, instead of the smooth cross-over predicted in the SM (for a recent discussion of the interplay between collider observables and models with a SFOPT, see e.g. Ref.~\cite{Ramsey-Musolf:2019lsf}). In the context of SM modifications of the Higgs properties~\cite{deFlorian:2016spz} parameterized by effective-field-theories (EFTs), it is well known that changes of the Higgs potential are often correlated with changes of other couplings, such as those of the Higgs to the EW gauge bosons. In many instances, a very precise measurement of the latter can be as powerful in constraining new physics as the self-coupling measurement~\cite{Katz:2014bha}. For example, Ref.~\cite{Huang:2016cjm} considered models for SFOPT with an extra real scalar singlet, and showed that a measurement of the HZZ coupling $g_{HZZ}$ with a precision of $\sim 1\%$ can rule out most of the parameter space that could be probed by a measurement of the self-coupling with a $\sim 50\%$ precision (see Fig.~1 of that paper). Should a deviation from the SM be observed in $g_{HZZ}$, however, a large degeneracy would be present in the set of allowed parameters. For example, Fig.~1 of Ref.~\cite{Huang:2016cjm} shows that a $\sim 2\%$ deviation in  $g_{HZZ}$ would be compatible, in this class of models, with any value of $1\lesssim \lthree/\lthree^{SM}\lesssim 2$. A precise direct measurement of $\lthree$ is therefore necessary, independently of what other observables could possibly probe, and is an indispensable component of the Higgs measurement programme.

Another remark is in order: the relation between the Higgs self-coupling and HH production properties is unambiguous only in the SM. Beyond the SM, the HH production rate could be modified not only by a change in the Higgs self-coupling, but also by the presence of BSM interactions affecting the HH production diagrams. These could range from a modified top Yukawa coupling, to higher-order EFT operators leading to local vertices such as ggHH~\cite{Azatov:2015oxa}, WWHH~\cite{Bishara:2016kjn} or \tthh~\cite{Grober:2010yv,Contino:2012xk}. The measurement of an anomalous HH production rate, therefore, could not be turned immediately into a shift of $\lthree$; rather, its interpretation should be made in the context of a complete set of measurements of both Higgs and EW observables, required to pin down and isolate the coefficients of the several operators that could contribute. In view of this, it is not possible to predict an absolute degree of precision that can be achieved on the measurement of $\lthree$, since this will depend on the ultimate $\lthree$ value, on the specific BSM framework leading to that value, and on the ancillary measurements that will be available as additional inputs. As is customary in the literature~\footnote{However, see for example Ref.~\cite{DiVita:2017vrr} and Refs.~\cite{Azatov:2015oxa,Capozi:2019xsi,Biekotter:2018jzu}, for global studies of the Higgs self-coupling in presence of multiple anomalous couplings, at $e^+e^-$ and pp colliders, respectively.}, we shall therefore focus on the context of the SM, neglecting the existence of  interactions influencing the HH production, except for the presence of a pure shift in $\lthree$. The precision with which $\lthree$ can be measured under these conditions has been for a long time the common standard by which the performance of future experiments is gauged, and we adopt here this perspective. Our results remain therefore indicative of the great potential of a hadron collider in the exploration of the Higgs potential.

\begin{figure}[ht!]
  \centering
    \includegraphics[width = 0.95\linewidth]{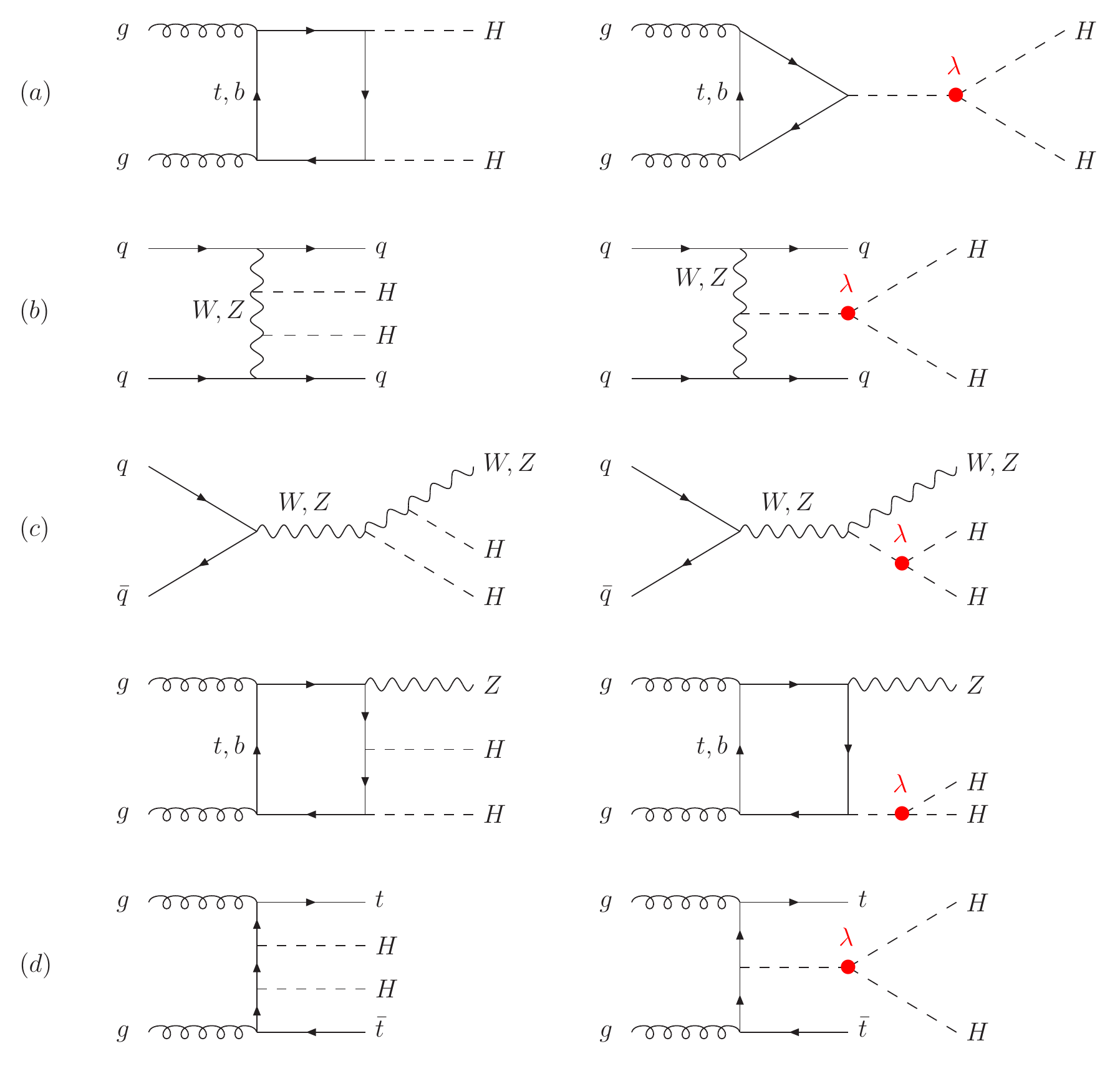}
    \caption{Diagrams contributing to Higgs pair production: (a) gluon fusion, (b) vector-boson fusion, (c) double Higgs-strahlung and (d) double Higgs bremsstrahlung off top quarks. The trilinear Higgs self-coupling is marked in red. }
  \label{fig:diagrams}
\end{figure}

\section{The theoretical modeling of signals and backgrounds}
\label{sec:simulation}
The signal and background processes are modeled with the  \MGAMC~\cite{Alwall:2014hca} and \pwg~\cite{Frixione:2007vw,Alioli:2010xd} Monte Carlo (MC) generators, using the parton distribution functions (PDF) set \pdfset~\cite{Ball:2014uwa} from the \lhapdf~\cite{Buckley:2014ana} repository. The evolution of the parton-level events is performed with \py~\cite{Sjostrand:2014zea}, including initial and final-state radiation (ISR, FSR), hadronization and underlying event (UE). The generated MC events are then interfaced with the \delphes~\cite{deFavereau:2013fsa} software to model the response of the FCC-hh detector, as described in  Section~\ref{subsec:reconstruction}. The full event generation chain is handled within the integrated FCC collaboration software (\fccsw)~\cite{fccsw_web}. The event yields for the background and signal samples are normalized to the integrated luminosity of \intlumifcc.


\subsection{The HH production processes}
\label{subsec:mcs}
At \sqrtsfcc, the dominant HH production modes are, in order of decreasing  cross section, gluon fusion (\gghh), vector boson fusion (\vbfhh), associated production with top pairs (\tthh) and double Higgs-strahlung (\vhh). A subset of diagrams for these processes is given in Fig.~\ref{fig:diagrams}.
Single top associated production is also a possible production mode but it is neglected in this study.
The cross-section calculations~\cite{Baglio:2012np,deFlorian:2013jea,Grigo:2014jma,deFlorian:2016uhr,Borowka:2016ypz,Dreyer:2018qbw,Grazzini:2018bsd,Baglio:2018lrj,Davies:2019dfy,Baglio:2020ini} for these main production mechanisms, reported also in Refs.~\cite{hhxswg,deFlorian:2016spz,DiMicco:2019ngk}, are given in Table~\ref{tab:xsecs_s}. We note that the relative rate of the sub-dominant modes (\vbfhh, \tthh and \vhh) increases significantly from \sqrtslhc\ to \sqrtsfcc. In particular, associated top pair production becomes as important as vector boson fusion, and together they contribute to nearly 15\% of the total HH cross section.

\begin{table}[ht!]
\renewcommand{\arraystretch}{1.5}
\begin{center}
\begin{tabular}{lcccc}
Process & $\sigma$(14 TeV) & $\sigma$(100 TeV) & accuracy & K-factor\\ \hline
\gghh\ & $36.69 \pm 5.3\%$ & $1224 \pm 5.6\%$ & $\mathrm{\nnlo_{FTapprox}}$ & 1.08 \\ 
\vbfhh\ & $2.05 \pm 2.1\%$ & $82.8 \pm 2.1\%$ & N$^3$LO & 1.15\\ 
\tthh\ & $0.949 \pm 2.9\%$ & $82.1 \pm 7.8\%$ & \nlo & 1.38\\ 
\vhh\ & $0.982 \pm 1.8\%$ & $16.23 \pm  2.9 \%$ & \nnlo & 1.40\\
\end{tabular} 
\caption{Signal cross sections ($\sigma$, in fb) for HH production, including the QCD corrections recommended by the LHC Higgs Cross Section Working Group \cite{hhxswg,deFlorian:2016spz}. For each process, scale variations have been symmetrized and added in quadrature to PDF+\as\ uncertainties. For the \gghh{} process, we added in quadrature also the dominant uncertainty induced by the finite $m_{top}$ corrections. The cross sections of \wmhh, \wphh\ and \zhh\ processes have been summed together in a single \vhh\ line and their uncertainties have been summed in quadrature.}
\label{tab:xsecs_s}
\end{center}
\end{table}

\begin{figure}[t!]
\centering     
  \subfigure[]
  {\label{fig:xsec_vs_lambda}
   \includegraphics[width = 0.47\linewidth]{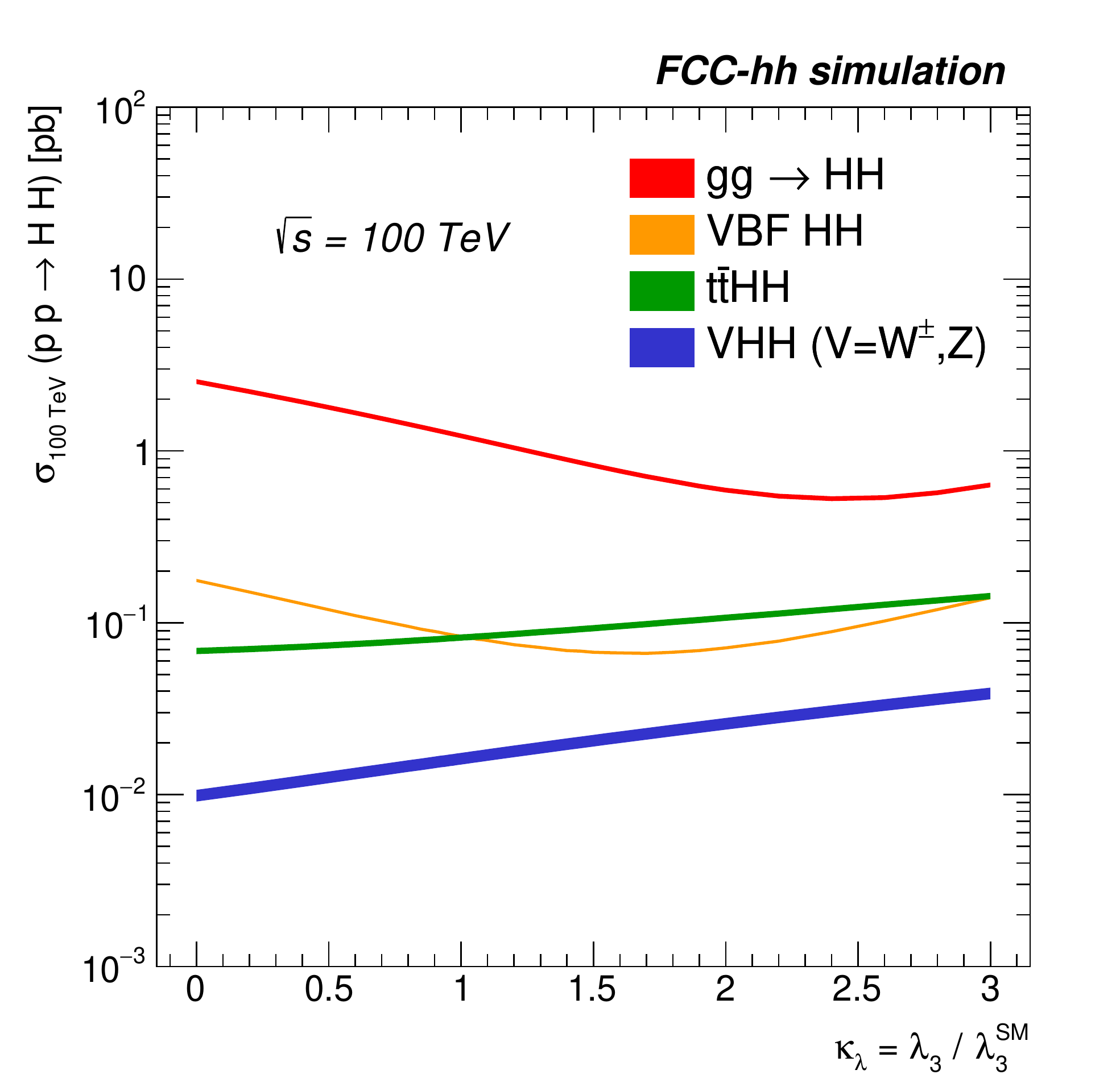}
  }
  \subfigure[]
  {\label{fig:hhptgen}
   \includegraphics[width = 0.47\linewidth]{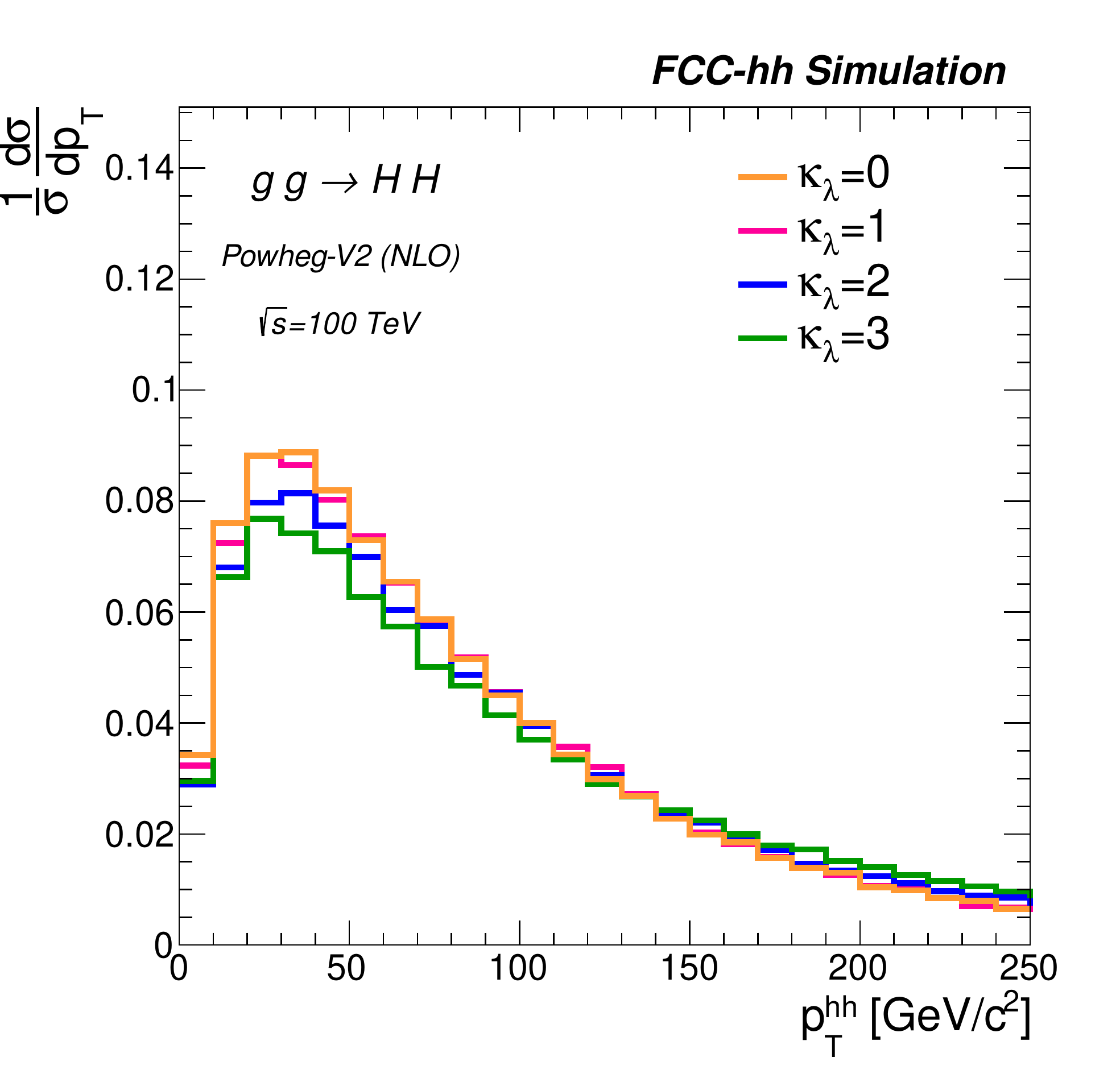}
  }
  \caption{(a) Cross section of the \gghh, \vbfhh, \tthh, and \vhh processes as a function of \klfrac. (b) Transverse momentum spectrum of the HH system in \gghh\ NLO events after parton-shower merging for $\kl=0$, $\kl=1$, $\kl=2$ and $\kl=3$.}
\end{figure}

The \gghh\ MC events have been generated at next-to-leading order (NLO) with the full top mass dependence using \pwg~\cite{Alioli:2010xd,Heinrich:2019bkc}. The \vbfhh, \tthh and \vhh\ events were instead generated at leading order (LO) with \MGAMC. All the HH production mechanisms feature the interference between diagrams that depend on the self-coupling with diagrams that do not. This leads to a non-trivial total cross section dependence on \lthree, as shown in Fig.~\ref{fig:xsec_vs_lambda}, and has crucial implications for the self-coupling measurement strategy, as discussed  in Section~\ref{subsec:genstrat}.
In order to account for this non-trivial dependence of the cross section on the self-coupling, the MC samples for the signal processes have been generated for several possible values of \klfrac\ within the interval $\kl \in$ [0.0,3.0]. In order to match the MC inclusive cross section prediction with the cross sections of Table~\ref{tab:xsecs_s}, we correct the event normalisation by means of a constant K-factor (shown in the last column of Table~\ref{tab:xsecs_s}). We note that in principle the K-factors are \kl-dependent. 
In this work, the (process dependent) K-factor is derived for $\kl=1$ and applied to correct the cross section at values of $\kl \neq 1$. This is justified by the explicit calculation of the N$^3$LO corrections at $\kl \neq 1$ for the VBF production channel~\cite{Dreyer:2018qbw}, and by the study of the $\kl$ dependence of the NNLO/NLO ratio for \gghh{} in Ref.\cite{Amoroso:2020lgh}. In the latter case, the shape variation of kinematical distributions for $\kl \neq 1$ from NLO to NNLO is small compared to the overall size of the difference between $\kl \neq 1$ and $\kl = 1$.
The total cross section obtained with this procedure as a function of \kl is shown in Fig.~\ref{fig:xsec_vs_lambda}. The merging of the NLO parton-level configurations with the parton-shower evolution is realized in the \pwg\ samples with \py. In Fig.~\ref{fig:hhptgen} the transverse momentum of the HH system \ptSup{HH}\ is shown as a validation of the NLO merging procedure. For the \vbfhh, \tthh{} and \vhh\ samples, \py\ simply adds the regular parton shower to the LO partonic final states.

The Higgs self-coupling can be probed via a number of different Higgs boson decay channels. Given the small cross section, at least one of the Higgs bosons is required to decay to a pair of b-quarks. Here, we consider the three most promising channels: \hhbbaa, \hhbbtata\ and \hhbbbb. The di-Higgs system decay in the various modes is performed by the \py\ program and the respective branching fractions $\br{\hhbbaa}=0.00262$, $\br{\hhbbtata}=0.072$ and $\br{\hhbbbb}=0.33$ are taken from Ref.~\cite{deFlorian:2016spz}, assuming $m_\mathrm{H}=125.10\,\GeV$.

\subsection{The background processes}
\label{subsec:mcb}

The background processes for the channels under study can be classified in irreducible, reducible and instrumental backgrounds. Irreducible backgrounds feature the presence in the matrix element of the exact same final state as the \gghh\ signal process. These include for example prompt \bbaa\ (QCD) production, or \zbb\ with \ztobbtata. We define as reducible background the processes that contain the same final state particles as the signal, but also additional particles that can be used as handles for discrimination. This is the case for instance of \tth, \haa as a background for the \hhbbaa\ channel or the \ttbar\ background for the \hhbbtata\ channel. Finally, we call as instrumental the background processes that mimic the signal final state due to a mis-reconstruction of the event in the detector. An instrumental background for the \hhbbaa\ channel is the \aj\ process where one the jets gets accidentally reconstructed as an isolated photon. Special care has to be given to such backgrounds as they strongly depend on the details of the detector performance.

Single-Higgs production constitutes a background for all di-Higgs final states. The four main production modes, gluon fusion (\ggh), vector boson fusion (\vbfh), top pair associated production (\tth) and Higgs-strahlung (\vh), have been simulated at LO, including up to two extra MLM-matched jets~\cite{Mangano:2006rw,Alwall:2007fs}, using \MGAMC . The \ggh\ matrix element was generated using the full top mass dependence. The rates of single-Higgs processes have been normalised to the most accurate cross-section calculations at \sqrtsfcc~\cite{Contino:2016spe}. The normalisation K-factor for the \ggh\ process includes corrections up to \nnnlo, while the \vbfh, \tth and \vh modes include corrections up to \nnlo.

Top-induced backgrounds, in particular top-pair production (\ttbar), constitute a large background for the \hhbbtata\ final state, and to a minor degree for the \hhbbbb\ final state. This process was generated at LO using \MGAMC\ with up to two extra MLM-matched jets. The total cross section is normalised to match the \nnlo\ prediction at \sqrtsfcc. The Drell-Yan (\zjets) and di-boson backgrounds are also mainly relevant for the \hhbbtata\ and \hhbbbb\ final states. These are generated at LO with \MGAMC\ by directly requiring the presence of \bbtata\ (or \jjbb\ for the \bbbb\ channel) final state at the matrix element level. The pure QCD contribution, \jjbb\, has also been generated with \MGAMC\ at order $\mathcal{O}(\as^3)$. The next contribution, \zjets, corresponding to \jjbb\ and \bbtata\ was generated at order $\mathcal{O}(\as^2 \aew)$. The latter includes for example the \ztobbtata\ process. The final contribution, generated at order $\mathcal{O}(\as \aew^2)$, includes the pure EW processes such as ZZ and ZH. When this background is included, the single-Higgs ZH mode discussed earlier is indeed omitted. For the pure QCD contribution we simply assume a conservative $K=2$ correction to the MC LO cross section. For the processes at orders $\mathcal{O}(\aew)$ and $\mathcal{O}(\aew^2)$ we employ K-factors that match to the \nnlo\ Drell-Yan and di-boson \sqrtsfcc\ predictions. The last class of relevant background processes for the the \hhbbbb\ and the \hhbbtata\ final states are the \ttz\ and \ttw\ processes. These were also generated at LO using \MGAMC\ and normalized to the highest accuracy NLO cross-section calculations.

The largest background contribution for the \hhbbaa\ final state are QCD multijet production with one or more prompt photons in the final states,  \aaj\ and \aj{} respectively. For the \aaj\ process we generated the matrix element of $\gamma \gamma$ plus two partons with \MGAMC, where partons are generated in the 5-flavour (5F) scheme to allow for mis-reconstructed light and c-quark jets. In order to maximise the MC event efficiency in the signal region, the \aaj\ process was generated with the $|\maa - 125| < 10$~GeV requirement at parton level. The \aj\ process was instead generated as $\gamma$ plus three partons in the final state, again in the 5F scheme. Both these processes were generated at LO and a conservative K=2 correction factor on the LO prediction to account for higher order predictions was applied. The \ttaa\ process was also considered for this channel and its contribution was found to be negligible.

\section{The experimental and analysis framework}
\label{sec:expsetup}
The FCC project is described in detail in its Conceptual Design Reports~\cite{Abada:2019zxq,Benedikt:2018csr}. We focus here on the 100~TeV pp collider, FCC-hh, designed to operate at instantaneous luminosities up to $\mathcal{L}=3\times 10^{35}~\ilum$. For our study we adopt the reference total integrated luminosity of \intlumifcc, to be achieved after 20 years of operations, possibly combining the statistics of two general purpose detectors. The analysis of these data will set challenging requirements to the detector design and performance, which will reflect on the physics potential in general, and in particular on the measurement of the HH cross sections.  We summarize here the main features of the current detector design, as implemented in the \delphes~\cite{deFavereau:2013fsa} simulation tool used for our study. 

\subsection{Detector requirements}
\label{subsec:detector}
A detector operating within the FCC-hh environment will have to isolate the hard-scattering event from up to 1000 pile-up (PU) simultaneous collisions per bunch-crossing. Extreme detector granularity together with high spatial and timing resolution are therefore needed. In addition, to meet the high precision goal in key physics channels such as \hhbbaa, an excellent photon energy resolution is needed. This requires a small calorimeter stochastic term~\footnote{The resolution in a calorimeter can be expressed as $\frac{\sigma_E}{E}=\frac{N}{E}\oplus\frac{S}{\sqrt{E}}\oplus C$, where $N$, $S$, and $C$ are usually referred respectively as the \emph{noise}, \emph{stochastic} and \emph{constant} terms.} in an environment of large PU noise, which in turn can be achieved via a large sampling fraction and a fine transverse and longitudinal segmentation. Finally, physics processes occurring at moderate energy scales ($Q=100\GeV - 1\TeV$) will be produced at larger rapidities compared to the LHC. Therefore high precision calorimetry and tracking need to be extended up to $\aeta<6$.

A prototype of a baseline FCC-hh detector that could fulfill the above requirements has been designed for the FCC CDR~\cite{Benedikt:2018csr,Aleksa:2019pvl,Selvaggi:2019xyd}. The detector has a diameter of 20\,m and a length of 50\,m, with dimensions comparable to the ATLAS detector. A central detector (covering a region up to $\aeta < 2.5$) contains a silicon-based tracker, a Liquid Argon (LAr) electromagnetic calorimeter (ECAL) and a Scintillating Tile Hadron calorimeter (HCAL) inside a 4\,T solenoid with a free bore diameter of 10\,m. The muon chambers are based on small Monitored Drift Tube technology (sMDTs). The tracking volume has a radius of 1.7\,m with the outermost layer lying at 1.6\,m from the interaction point (IP) in the central and the forward regions, providing the full lever arm up to $\aeta = 3$. The ECAL has a thickness of 30 radiation lengths and provides, together with the HCAL, an overall calorimeter thickness of than 10.5 nuclear interaction lengths. The transverse segmentation of both the electromagnetic and hadronic calorimeters is $\sim4$ times finer than the present ATLAS~\cite{Aad:2008zzm} and CMS calorimeters~\cite{Bayatian:2006nff}.
A high longitudinal segmentation in the ECAL is needed to ensure a high sampling fraction, hence a small stochastic term and in turn the good photon energy resolution required in order to maximise the efficiency of the \haa{} reconstruction. 
In order to reach good performances at large rapidites ($2.5 < \aeta < 6$), the forward parts of the detector are placed at 10\,m from the interaction point along the beam axis. Two forward solenoids with an inner bore of 5\,m provide the required bending power for forward tracking. The integrated forward calorimeter system (ECAL and HCAL) is fully based on LAr due to its instrinsic radiation hardness. Coverage up to $\aeta = 6$ is feasible by placing the forward system at a distance z$=$16.6\,m from the IP in the beam direction and at r$=$8\,cm in the transverse direction. The FCC-hh baseline detector performance has been studied in full \geant~\cite{Agostinelli:2002hh} simulations and parameterised within the fast simulation framework \delphes~\cite{deFavereau:2013fsa, delphes_card_fcc}.

\subsection{Detector simulation and object reconstruction}
\label{subsec:reconstruction}
The reconstruction of the MC-generated events in the FCC-hh detector is simulated with the \delphes\ framework. \delphes\ makes use of a parameterised detector response in the form of resolution functions and efficiencies. The \delphes\ simulation includes a track propagation system embedded in a magnetic field, electromagnetic and hadron calorimeters, and a muon identification system. \delphes\ produces physics objects such as tracks, calorimeter deposits and high level objects such as isolated leptons, jets, and missing energy. \delphes\ also includes a particle-flow reconstruction that combines tracking and calorimeter information to form particle-flow candidates, i.e charged hadrons, neutral hadrons and photons. Such particles are then used as input for jet clustering, missing energy, and isolation variables. In the following we will focus on describing the key parameters of the FCC-hh detector implementation in \delphes\ that are relevant for the self-coupling analysis presented here.

Jets are clustered by the \ak\ algorithm~\cite{Cacciari:2008gp} with a parameter R$=$0.4. For leptons ($\ell=e,\mu$) and photons ($\gamma$), the relative isolation \reliso\ is computed by summing the \pt\ of all particle-flow candidates in a cone around the particle of interest an dividing by the particle's $\pt(e,\mu,\gamma)$. Isolated objects, such as photons originating from a \hhbbaa\ decay, typically feature a small value of \reliso. The reconstruction and identification (ID) efficiencies for leptons and photons are parameterised as function of \pt\ and pseudo-rapidity $\eta$. 


We note that the effect of pile-up is not simulated directly by overlaying minimum bias events to the hard scattering. Although \delphes\ allows for such possibility, including in the simulation up to 1000 pile-up interactions would result in an overly conservative object reconstruction performance for the simple reason that the current \delphes\ FCC-hh setup does not possess the well-calibrated pile-up rejection tools that will necessarily be employed for a detector operating in such conditions, and so far in the future. These techniques will include the use of picosecond (ps) timing detectors as well as advanced machine-learning-based techniques for pile-up mitigation. For the present LHC detectors, as well as for presently approved future detectors (the ATLAS and CMS Phase II detectors) it is already the case that such techniques allow to recover the nominal detector performance in the absence of pile-up~\cite{Contardo:2020886, CERN-LHCC-2015-020}. The level of degradation of the \lthree\ measurement precision caused by the deterioration of the performance of specific physics objects (for example the photon energy resolution and reconstruction efficiency or the b-tagging efficiency) has been quantified in previous studies~\cite{L.Borgonovi:2642471}. The impact of degrading the photon energy resolution due to pile-up contamination was studied in full simulation with up 1000 pile-up interactions in Ref~\cite{Aleksa:2019pvl}. The degraded resolution was then propagated in \delphes\ and the effect on the \lthree\ precision was found to be approximately 1\% (or 20\% in relative terms). We stress however that this level of degradation should be considered as a worse case scenario given that a simple sliding window algorithm was used, and timing information was not exploited. In Ref.~\cite{L.Borgonovi:2642471} we have also studied the impact of degrading the photon reconstruction or, equivalently, of increasing the jet-to-photon probability, which also showed an effect of 1\% on the \lthree\ precision. 
Since a full-fledged event simulation and object reconstruction does not exist at this stage for the FCC-hh detector, the assumed object efficiencies result from extrapolations from the LHC detectors. In order to account for a possible degradation of the detector performance in the presence of pile-up, we define 3 baseline scenarios:

\begin{itemize}
    \item scenario I: optimistic -- target detector performance, similar to Run~2 LHC conditions
    \item scenario II: realistic -- intermediate detector  performance 
    \item scenario III:  conservative -- pessimistic detector performance, assuming extrapolated HL-LHC performance using present-day algorithms  
\end{itemize}

The assumptions on the performance of various physics objects for each baseline scenario, are summarized in Table~\ref{tab:detperf}. As mentioned previously, a dominant background for \hhbbaa\ analysis is the \aj\ process. The probability for a hard scattering jet to be mis-reconstructed as an isolated photon is small, $\mathcal{O}(10^{-3})$, in current LHC detectors, thanks to the excellent angular resolution of present calorimeters. As we noted in Section~\ref{subsec:detector}, the assumed granularity for the FCC-hh detector is a factor 2-4 better than present LHC detectors. We make however the conservative choice of assuming a $\text{j}\rightarrow\gamma$ fake-rate $\misg = 0.0007 \cdot e^{-\pt [GeV]/187}$, which is of the same magnitude as in the HL-LHC detectors~\cite{ATLAS:2016ukn}. In addition, we account for the probability for a pile-up jet to be reconstructed as photon by further multiplying the fake rate by a factor 2. This factor has been derived by simulating 1000 PU collision with the CMS Phase-II detector in Delphes, applying a pile-up ID mistag rate of 10~\% (from Ref.~\cite{CERN-LHCC-2015-020}) and applying the fake-rate probability for calibrated pile-up jets given in Figure 5(b) in Ref.~\cite{ATLAS:2016ukn}. This procedure is used for Scenario I. For Scenario II and III we multiply the above fake-rate by factors of 2 and 4 respectively. 
For leptons we neglect possible fake jets contributions since these are negligible at the momemtum scale relevant for the \hhbbtata\ final state. \delphes\ also provides heavy flavour tagging, in particular $\tau$ (hadronic) and $b$-jet identification.
Both hadronic $\tau$'s and $b$-jets are reconstructed using the total visible 4-momentum of the jet. The tagging efficiencies rely on a parameterisation of the (mis-)identification probability as a function of (\pt, $\eta$). Again, since we cannot yet derive such performance from full-simulation, we assume efficiencies and mistag rates of the same order as for the (HL-)LHC detectors. For Scenario I, the efficiencies for $\tau$ and $b$-jets are modelled after the CMS performance given in Refs.~\cite{CMS-DP-2019-033,CMS-DP-2017-013}. For scenarios II and III a degradation of the efficiencies for a constant mistag-rate probability is assumed. It should be noted that this is a conservative assumption, since present estimates for heavy-flavour tagging in LHC Phase II conditions project a similar performance as in present conditions, due to superior tracking and high-precision timing detectors~\cite{CMS:2667167,CERN-LHCC-2017-009}. For $\tau$ and $b$-jets we also consider two definitions, a "Medium" (M) and a "Tight" (T) working point, in order to operate at an optimal signal-to-background rejection in each decay channel. As shown in Table~\ref{tab:detperf}, we also consider the impact of photon and b-jet energy resolutions. Both are relevant since all di-Higgs processes are resonant and the reconstructed Higgs mass directly affects the final sensitivity. The di-photon resolution for scenarios I, II and III was directly determined from full-simulation respectively with 0, 200, and 1000 pile-up interactions (see Ref.~\cite{Aleksa:2019pvl}). For the di-jet invariant mass, for scenario I we assume the invariant mass resolution as obtained with a multi-variate regression technique in CMS Run~2 (in Ref.~\cite{Sirunyan:2018kst}), whereas for scenario II and III respectively we assume a factor 1.5 and 2 degradation compared to scenario I.  The \bbbb\ channel is a special case, since two di-jet invariant masses are reconstructed (see Section~\ref{subsec:bbbb}). In that case, as a conservative assumption, we assume that only the Higgs candidate with the largest \pt\ is affected by the \mbb\ resolution assumed for the scenarii I to III, while the sub-leading Higgs candidate mass has the resolution of Scenario III.  

\begin{table}[ht!]
\centering
\begin{tabular}{lccc}

parameterisation & scenario I & scenario II & scenario III    \\
\hline
\hline

b-jet ID eff.   &  82-65\% & 80-63\%  &  78-60\%   \\
b-jet c mistag   &   15-3\% & 15-3\%   & 15-3\% \\ 
b-jet l mistag  &   1-0.1\% & 1-0.1\%   & 1-0.1\%   \\ 
\hline
$\tau$-jet ID eff  &  80-70\% & 78-67\%  &  75-65\%    \\
$\tau$-jet mistag (jet)  & 2-1\% & 2-1\%   &   2-1\%   \\ 
$\tau$-jet mistag (ele) &  0.1-0.04\% & 0.1-0.04\%   &  0.1-0.04\%  \\ 
\hline      
$\gamma$ ID eff.  &  90 & 90  & 90   \\ 
jet $\rightarrow$ $\gamma$ eff.  &  0.1 & 0.2 & 0.4 \\
\hline
$m_{\gamma\gamma}$ resolution [GeV] &  1.2 & 1.8 & 2.9 \\ 
\hline
$m_{bb}$ resolution  [GeV] & 10 & 15 & 20  \\ 

\end{tabular}
\caption{Performance of physics objects for the various scenarios. Objects efficiencies and mistag rates are given for a representative $\pt\approx50\GeV$. For b and $\tau$-tagging (and their respective mistag rates) numbers for two different working points are given (Medium and Tight).}
\label{tab:detperf}
\end{table}


\subsection{Systematic uncertainties}
\label{subsec:systematics}

Systematic uncertainties can play a major role on the expected sensitivity of the self-coupling measurement. Several assumptions have been made on the possible evolution of theoretical and experimental sources of uncertainties in order to present a realistic estimate of the physics potential of FCC-hh for the channels considered here. In particular, for each uncertainty source, we defined three possible scenarios, following the general principle introduced in Section~\ref{subsec:reconstruction}. We note that the intermediate assumptions are almost equivalent to those made for HL-LHC projections~\cite{Cepeda:2019klc, deBlas:2019rxi}. 

\begin{table}[ht!]
\centering
\begin{tabular}{lcccc}

Uncertainty source & scenario I & scenario II & scenario III & Processes \\ \hline
b-jet ID eff. /b-jet & 0.5\% & 1\%  & 2\% & single H, HH, ZZ  \\ 
$\tau$-jet ID eff. /$\tau$ & 1\% & 2.5\% & 5\% & single H, HH, ZZ \\ 
$\gamma$ ID eff. /$\gamma$  & 0.5\%& 1\% & 2\% & single H, HH  \\ 
$\ell$ = $e$-$\mu$ ID efficiency & 0.5\% & 1\% & 2\% & single H, HH, ZZ\\ 
luminosity & 0.5\% & 1\% & 2\% & single H, HH, ZZ \\ 
theoretical cross section & 0.5\% & 1\% & 1.5\% & single H, HH, ZZ \\ 

\end{tabular}
\caption{Summary of the sources of systematic uncertainties in the 3 scenarios. The last column indicates the processes that are affected by the corresponding source of uncertainty. For each given object (b-jet, $\tau$-jet, $\gamma$, lepton), the quoted uncertainty on reconstruction and identification efficiency is applied as many times as the object appears in the final-state.}
\label{tab:syst}
\end{table}

A detailed list of the systematic uncertainties considered is presented in Table~\ref{tab:syst} for all the channels, together with the processes affected by each uncertainty. The numbers in the table refer to the individual contributions to the overall yield uncertainty. In particular, we consider uncertainties on:
\begin{itemize}
    \item \textbf{theoretical cross-section}, affecting the single-Higgs and ZZ backgrounds. Due to their moderate yields we assume these backgrounds to be estimated from Monte Carlo at the FCC-hh. We also assume these two processes to be well known and well reproduced by Monte Carlo simulations at the FCC-hh, with an overall uncertainty varied between 0.5\% and 1.5\% depending on the scenario. Furthermore, we include a similar theoretical uncertainty on the HH cross section, affecting the interpretation of the HH rate measurement in terms of $\mu$ and \kl.
    
    \item \textbf{luminosity}. We assume that the integrated luminosity will be known at FCC-hh at least as well as at the LHC. For this reason, we assume a conservative estimate of 2\% and an optimistic (intermediate) estimate of 0.5\% (1\%), reflecting future opportunities to extract the luminosity from hard processes like Z production. As for the theoretical uncertainties, the luminosity affects both the signal and the single-Higgs and ZZ backgrounds. 
    
    \item \textbf{experimental uncertainties} on objects reconstruction and identification efficiencies:
    \begin{itemize}
        \item[--] \textbf{b-jets}: for each b-jet, we assume a 0.5\%, 1.0\%, and 2.0\% uncertainty for the optimistic, intermediate and conservative scenarios, respectively. Since we expect this to be one of the dominant uncertainties, it is applied during event simulation accounting for the \pt dependence of the b-jet efficiency uncertainty (taken from Ref.~\cite{Sirunyan:2017ezt}). This procedure allows to take into account the effect of the uncertainty on the shape of the resulting BDT distribution.
        \item[--] \textbf{$\tau$-jets}: for each jet originated from the hadronic decay of a $\tau$-jet we assume an uncertainty of 1.0\%, 2.0\% and 5.0\% for the optimistic, intermediate and conservative scenarios, respectively.
        \item[--] \textbf{leptons}: we assume the same uncertainty on the lepton identification and reconstruction efficiency for electrons and muons: a 0.5\%, 1.0\%, and 1.5\% uncertainty for the optimistic, intermediate and conservative scenarios, respectively. 
        \item[--] \textbf{photons}: we assume that the photon related uncertainties will be comparable to electrons. For this reason we assign a systematic uncertainty to photon reconstruction of 0.5\%, 1.0\%, and 2.0\% for the optimistic, intermediate and conservative scenarios, respectively.
    \end{itemize}
\end{itemize}

The uncertainty on the energy resolution of photons has been studied in Ref.~\cite{L.Borgonovi:2642471}. Degrading the photon resolution by a relative 100\% has an effect of order 1\% on the self-coupling precision. According to Ref.~\cite{Aad:2014nim}, the uncertainty on the resolution, parametrised as an additional constant term, amounts to 0.4\% in the barrel. Assuming a photon energy E$=$60~GeV, a nominal constant term of 0.8\% and a stochastic term of 10\%, this results in a relative difference of less than 5\% (in relative terms~\footnote{More precisely the relative energy resolution is 1.57\% with the additional 0.4\% constant term and 1.51\% without the additional 0.4\% constant term}
) on the photon energy resolution. Such a degradation has an effect, at first order, of less than 0.1\% of the self-coupling measurement (assuming that a degradation of more than 100\% has an effect of 1\% on the self-coupling precision). A similar argument applies to the effect of the jet energy resolution, which is measured by ATLAS with a relative precision of 7\% according to Ref.~\cite{Aaboud:2019ibw}. We therefore neglect such source of systematic uncertainties. As far as the scale uncertainty goes, for both photons and jets, the above references show that they amount respectively to 0.1\% and 1\%. We have then explicitly verified that the effect of shifting the scale of both photon and jets by 1\% results in a relative change in the significance of order 0.5\%, translating at worst into a relative change in the self-coupling precision~\footnote{We assume a conservative factor 2 between the precision on the self-coupling and the precision on the self-coupling. See Section~\ref{subsec:genstrat} for a discussion on how the two are related} of 1\%. We conclude therefore that also the photon and jet (and necessarily hadronic $\tau$'s) energy scale have a negligible impact on the self-coupling precision. Moreover, we stress that the effect of a degradation of the detector performance, especially in terms of photons and b-jet energy resolution, is probed by studying the various scenarios given in Table~\ref{tab:detperf}. The absolute performance degradation considered in Scenario II and III largely overcomes the corresponding systematic uncertainties on the object resolution that were neglected.

We assume that several backgrounds will be measured with high statistical accuracy from ``side bands'' or ``control regions''. This is the case for example for the \ttbar, QCD, and non-single-Higgs backgrounds (with the exception of ZZ, that we assume to be predicted by the Monte Carlo) that dominate the \bbbb\ and \bbaa\ channels background contributions. In these cases, while there is no uncertainty associated to the normalisation of the backgrounds, the statistical uncertainty due to the possible fluctuations of the number of events in the side bands is considered in the fit. 

When performing the fit for the combination across different channels, systematic uncertainties with the same physical origin are considered fully correlated across processes and final states. Otherwise they are considered as completely uncorrelated, with the notable exception of the b-tagging efficiency (and mistag rate) uncertainty, which is correlated across channels but not across processes. We therefore consider  separate shape b-tagging and mistag uncertainties for each process (single H, ZZ, and HH), correlated across the various channels. For example, the b-tagging uncertainty affecting the single-Higgs production is correlated across all channels, but is uncorrelated from the the b-tagging uncertainty affecting double Higgs, ZZ, and so on. This reflects possible differences in the properties of b-tagged jets created in different processes.

%


\section{Signal extraction methodology}
\label{sec:strategy}
\label{subsec:genstrat}

As mentioned in Section~\ref{subsec:mcs}, the cross section for HH production has a non-trivial dependence on the self-coupling modifier \klfrac\ due to the presence, at LO, of diagrams that contain the trilinear interaction vertex (\sd) as well as diagrams that do not (\td), as shown in Fig.~\ref{fig:diagrams}. In Fig.~\ref{fig:diagrams}, \td-diagrams appear on the left column while \sd-diagrams are shown on the right. The \sd\ and \td\ contributions are present in all HH-production mechanisms. Moreover, the contribution of the interference term between \sd\ and \td\ is highly non trivial. For the \gghh\ and \vbfhh\ modes, the total cross section reaches a minimum respectively at $\kl\approx$~2.5 and $\kl\approx$~1.8, while the slopes of the \tthh\ and \vhh\ cross sections carry little dependence on \kl (see Fig.~\ref{fig:xsec_vs_lambda}).

At first order one can write:
\begin{equation}
\mu(\kl) = 1 + (\kl-1)\slope,
\label{eq:mu_vs_kl}
\end{equation}
where we define \mufrac\ as the signal strength. One can measure \lthree (or alternatively \kl) by measuring the total HH production cross section. It follows that:
\begin{equation}
\dkl = \frac{\dmu}{\slope},
\label{eq:err_kl}
\end{equation}
where \dkl\ and \dmu\ are respectively the uncertainty on the self-coupling modifier and on the signal strength. It can be noted that at first order the precision of the self-coupling measurement is determined by the slope of the cross section (or $\mu$) at $\kl=1$ and by the uncertainty on the measurement of the total cross section. Since \slope\ is a given parameter, in order to maximise the precision on the self-coupling, we have to maximise the precision on the cross section, or equivalently on $\mu$. Assuming all other standard model parameters are known with better precision than the expected precision on \kl~\footnote{This assumes that for instance the top Yukawa parameter will be known with \reldyt $\approx$ 1\%. The studies of Refs.~\cite{Plehn:2015cta,L.Borgonovi:2642471} show that such precision is achievable at the FCC-hh, using the ttZ coupling measured at FCC-ee~\cite{Janot:2015yza}}, the relative weight of the \sd\ and \td\ amplitudes (and their interference) is determined by the magnitude of \kl.

\begin{figure}[t!]
\centering     
  \subfigure[]
  {\label{fig:gghh_hhm}
   \includegraphics[width = 0.47\linewidth]{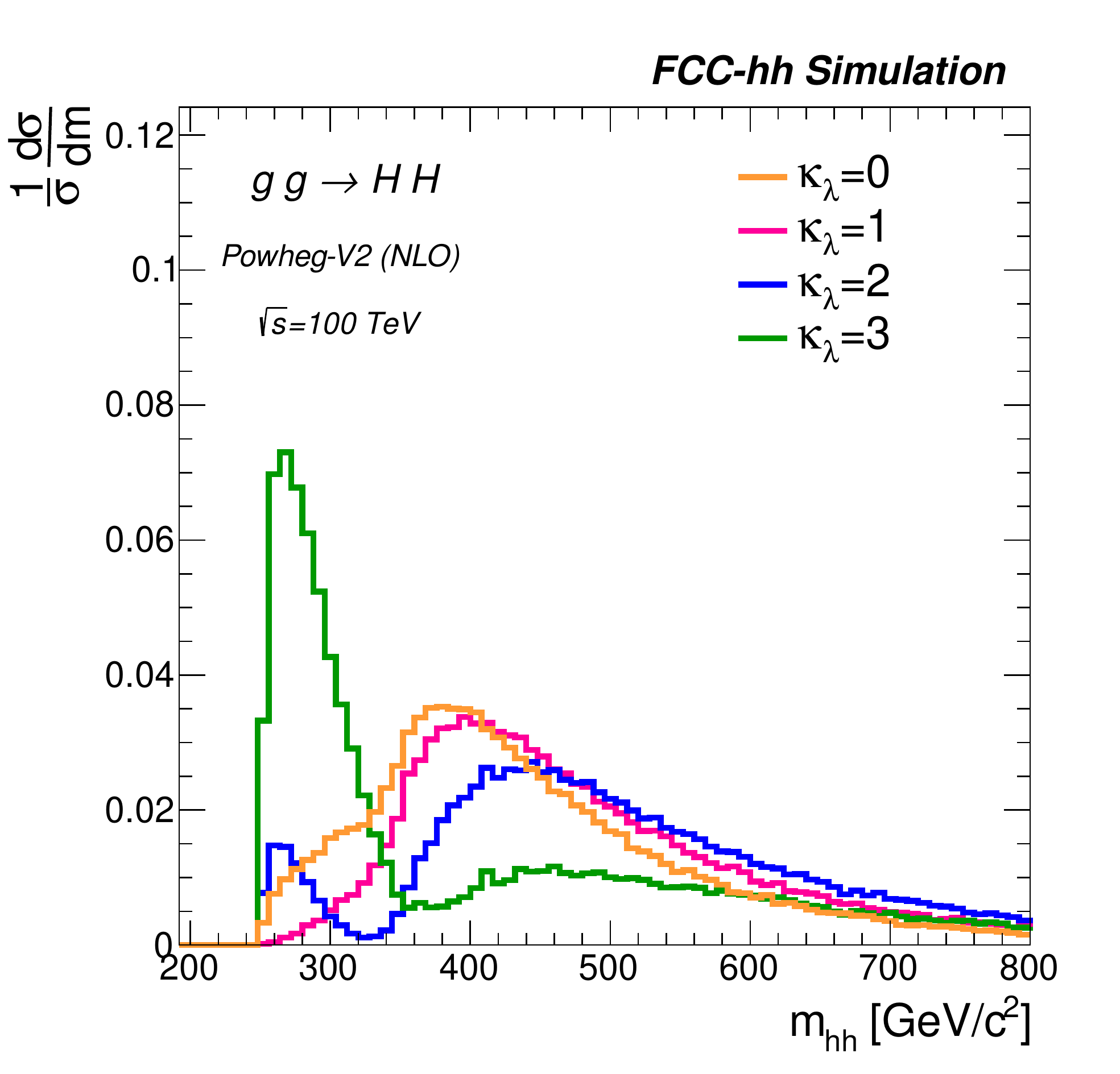}
  }
  \subfigure[]
  {\label{fig:tthh_hhm}
   \includegraphics[width = 0.47\linewidth]{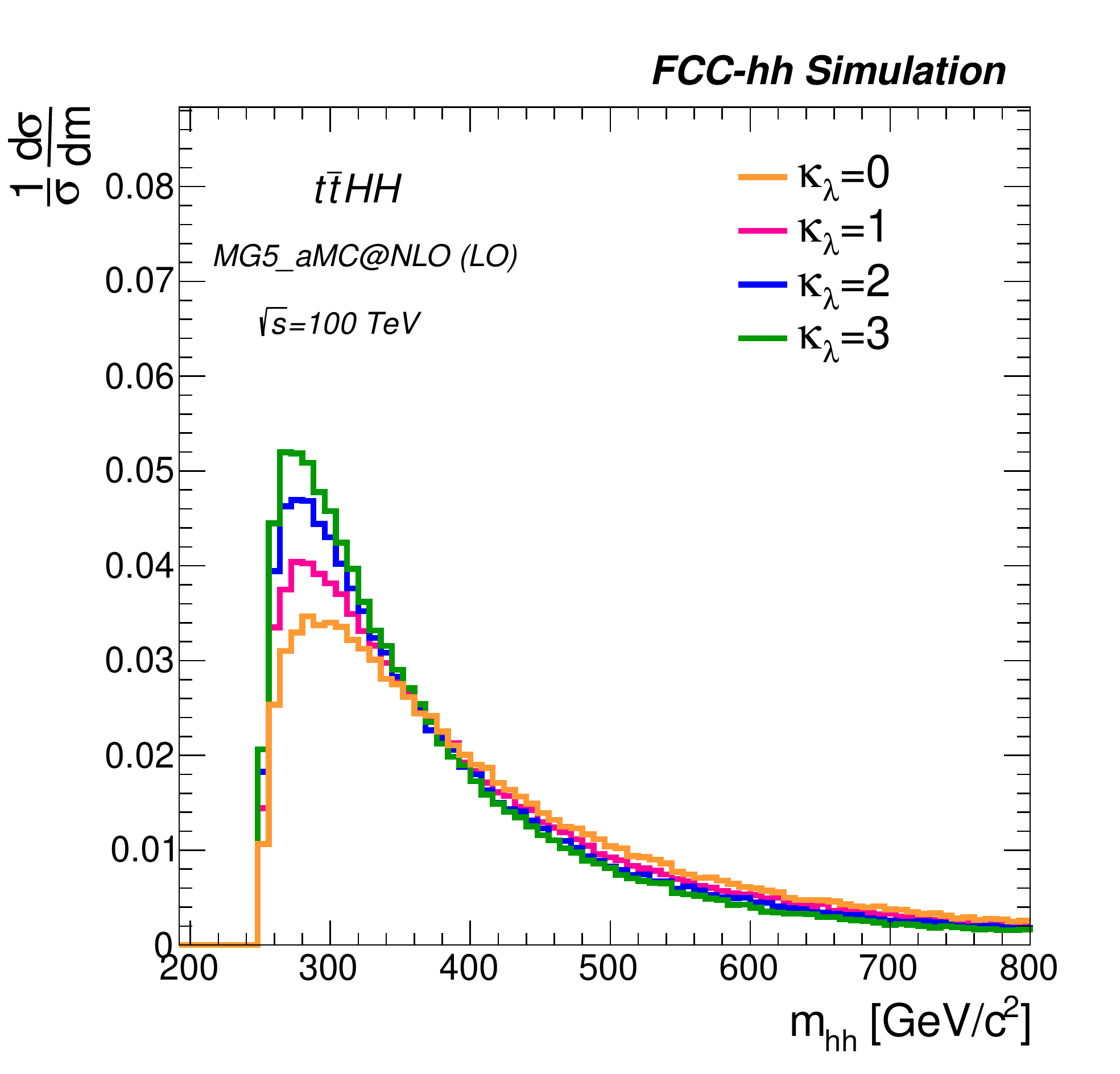}
  }
  \caption{Higgs pair invariant-mass distribution in \gghh\ (a) and \tthh (b) events for $\kl=0$, $\kl=1$, $\kl=2$ and $\kl=3$.}
\end{figure}

\begin{figure}[ht!]
\centering     
  \subfigure[]
  {\label{fig:gghh_h1pt}
   \includegraphics[width = 0.47\linewidth]{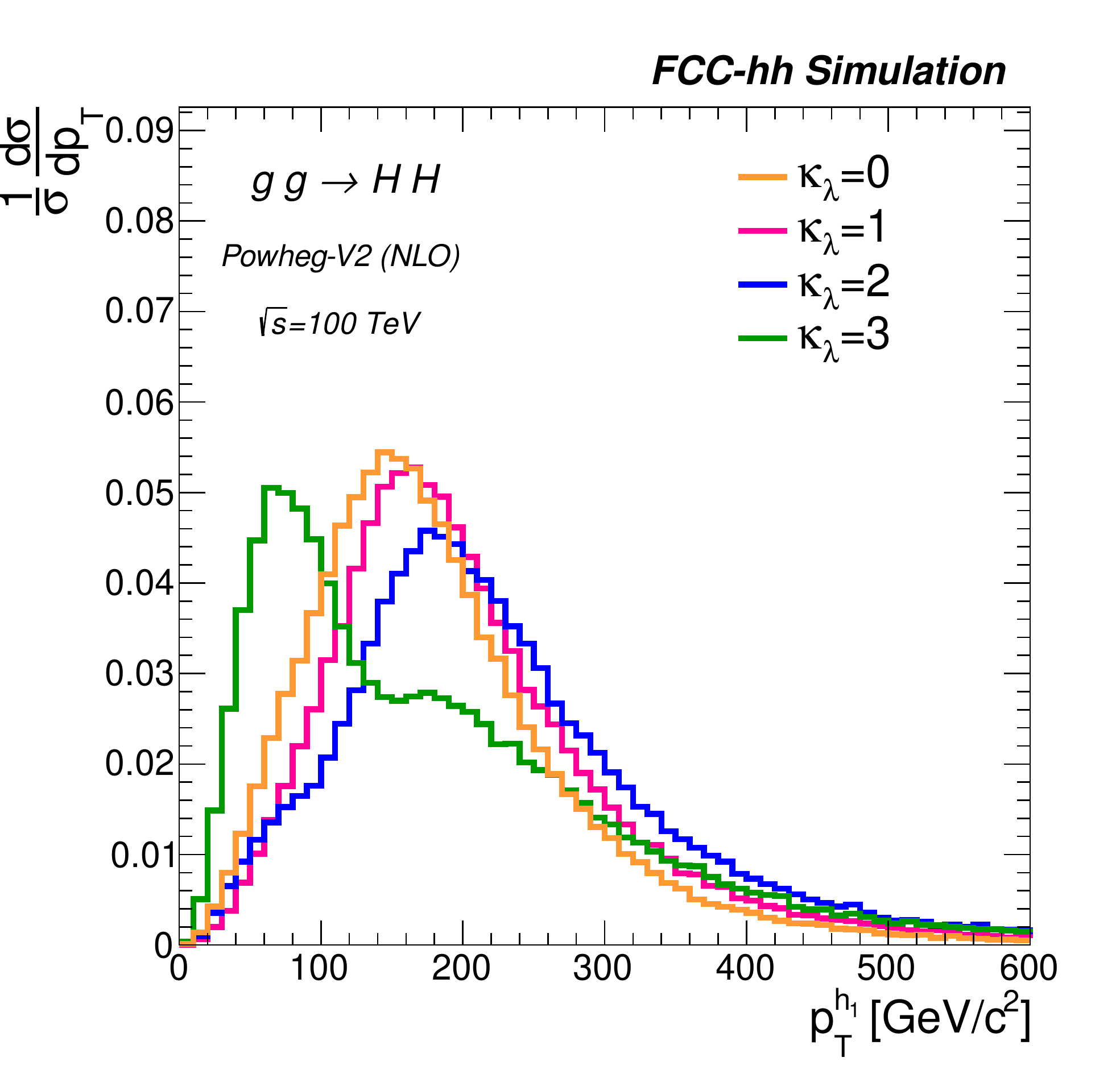}
  }
  \subfigure[]
  {\label{fig:gghh_h2pt}
   \includegraphics[width = 0.47\linewidth]{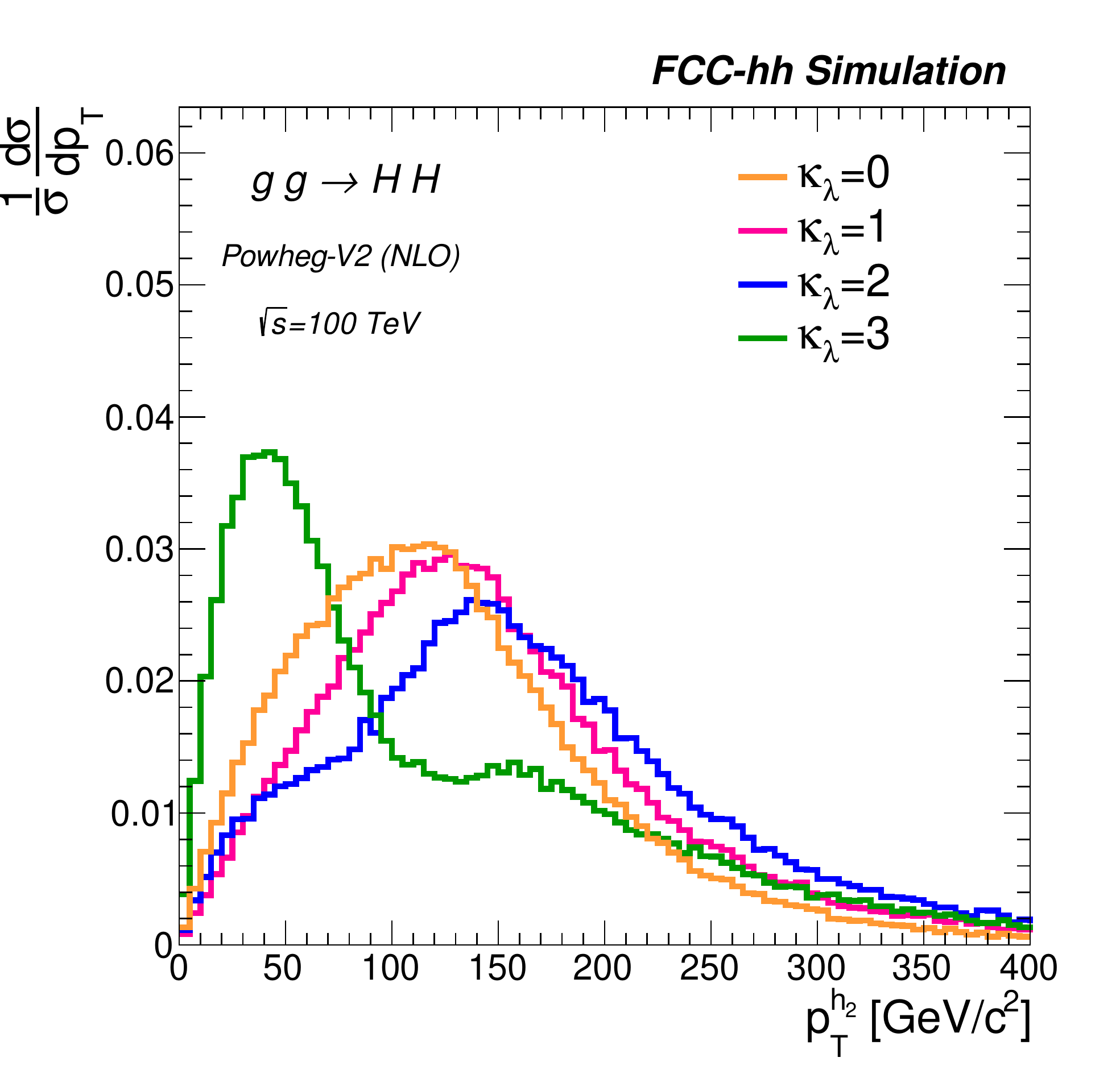}
  }
  \caption{Transverse momentum spectra of the leading (a) and sub-leading (b) Higgs boson in \gghh\ events for $\kl=0$, $\kl=1$, $\kl=2$ and $\kl=3$.}
\end{figure}

The magnitude of \kl\ impacts not only the total HH rate, as discussed above, but also the HH production kinematic observables. Notably, the invariant mass of the HH pair \mhh\ is highly sensitive to the value of the self-coupling. This can be easily understood by noting that configurations with large \mhh\ are mostly suppressed in the \sd\ amplitude (not in \td\ diagrams). Vice-versa, the phase-space region near threshold, at \mhh\ $\gtrsim$ 2\mH, maximises the \sd\ contribution.  
The \mhh\ distribution is shown for the \gghh\ and \tthh\ processes respectively in Figs.~\ref{fig:gghh_hhm} and~\ref{fig:tthh_hhm}. For \gghh\ the dependence is distorted at values of \kl\ $\approx$ 2 due to the large destructive interference between \sd\ and \td. The transverse momenta of the two Higgs bosons (\pthmax, and \pthmin) also display a large dependence on \kl, as shown in Figs.~\ref{fig:gghh_h1pt} and~\ref{fig:gghh_h2pt}.

The general strategy for providing the best possible accuracy on the self-coupling will therefore rely on maximizing the cross-section precision by using obervables that are able to discriminate between signal and backgrounds as well as exploiting the shapes of observables that are highly sensitive to the value of \kl. The signal over background optimisation is largely dependent on the class of background and will be addressed in the discussion specific to each channel below. However a common theme is that typically the strategy to obtain a high \sob\ ratio relies heavily on the reconstruction of the mass peak of the two Higgs bosons. In addition we will make use of the \mhh\ observable, and the Higgs particles transverse momentum (\pthmax, and \pthmin) differential distributions to further improve the sensitivity on \kl.

\section{Determination of the Higgs self-coupling}
\label{sec:analysis}
While the Higgs pair can be reconstructed in a large variety of final states, only the most promising ones are considered here: \bbaa, \bbtata\ and \bbbb. For each of these final states, the event kinematical properties are combined within boosted decision trees (BDTs) to form a powerful single observable that optimally discriminates between signal and backgrounds. The BDT discriminant is built using the ROOT-TMVA package~\cite{Brun:1997pa,Hocker:2007ht}. The statistical procedure and the evaluation of the systematic uncertainties are summarized in Appendix~\ref{subsec:procedure}  and~\ref{subsec:systematics}, respectively.

For a similar analysis in the case of HL-LHC, see Ref.~\cite{Adhikary:2017jtu} and the studies by the ATLAS and CMS collaborations~\cite{ATL-PHYS-PUB-2018-053,CMS-PAS-FTR-18-019}, contributed to Ref.~\cite{Cepeda:2019klc}.

\subsection{The \bbaa\ channel}
\label{subsec:bbaa}

Despite its small branching fraction, the \hhbbaa\ channel is by far the most sensitive decay mode for measuring the Higgs self-coupling. The presence of two high \pt\ photons in the final state, together with the possibility of reconstructing the decay products of both Higgses without ambiguities and with high resolution, provide a clean signature with a large \sob. The largest background processes are single-Higgs production and the QCD continuum \aaj\ and \aj. A discussion of the simulation of these processes
was given in Section~\ref{subsec:mcb}.

\subsubsection{Event selection}

In the \bbaa\ channel, events are required to contain at least two isolated photons and two b-tagged jets with the requirement \ptab~>~30~GeV and \etaab~<~4.0. The leading photon and b-jet are further required to have \ptab~>~35~GeV. The Higgs candidates 4-momenta are formed respectively from the two reconstructed b-jets and photons with the largest \ptab. The b-jets are identified with the "Medium" working point criterion, defined in Table~\ref{tab:detperf}. Since the \aaj\ process was generated  with a parton-level requirement (see Section~\ref{subsec:mcb}) on \maa, we further require the events to pass the loose selection $|\maa - 125| < 7$~GeV. The efficiency of the full event selection for the SM signal sample is approximately 26\%. For an integrated luminosity \intlumifcc\, this event selection yields approximately 10k Higgs pair events, 125k single Higgs, 2.6M \jjaa\ and 7M \aj events for Scenario I. The trigger efficiency for the above selection is assumed to be 100\% efficient.

In order to maximally exploit the kinematic differences between signal and background, a boosted decision tree (BDT) is trained using most of the available kinematic information in the event:

\begin{itemize}
      \item The 3-vector components of the leading (\amax) and subleading photon (\amin): transverse momentum (\ptSup{\amax}, \ptSup{\amin}), pseudo-rapidity (\etaSub{\amax}, \etaSub{\amin}), and azimutal angle (\phiSub{\amax}, \phiSub{\amin}).

  \item The 3-vector components of the leading (\bmax) and subleading b-jet (\bmin): transverse momentum (\ptSup{\bmax}, \ptSup{\bmin}), pseudo-rapidity (\etaSub{\bmax}, \etaSub{\bmin}), and azimutal angle (\phiSub{\bmax}, \phiSub{\bmin}). 
  
  \item The 3-vector components of the leading (\jmax) and subleading additional reconstructed jets in the event (\jmin): transverse momentum (\ptSup{\jmax}, \ptSup{\jmin}), pseudo-rapidity (\etaSub{\jmax}, \etaSub{\jmin}), and azimutal angle (\phiSub{\jmax}, \phiSub{\jmin}). If no additional jets are found, dummy values are given to these variables. 

  \item The 4-vector components of the \haa\ candidate: transverse momentum (\ptSup{\aa}), pseudo-rapidity (\etaSub{\aa}), azimutal angle (\phiSub{\aa}) and invariant mass (\maa).

  \item The 4-vector components of the \hbb\ candidate: transverse momentum (\ptSup{\bb}), pseudo-rapidity (\etaSub{\bb}), azimutal angle (\phiSub{\bb}) and invariant mass (\mbb).

  \item The 4-vector components of the Higgs pair candidate: transverse momentum (\ptSup{\hh}), pseudo-rapidity (\etaSub{\hh}), azimutal angle (\phiSub{\hh}) and invariant mass (\mhh).

  \item The number of reconstructed b-jets $N_b$, the number of light jets $N_l$ and the total number of jets $N_j = N_b + N_l$.
  
\end{itemize}

\begin{figure}
\centering     
  \subfigure
  {\label{fig:haa_m}
   \includegraphics[width = 0.31\linewidth,bb = 10 0 525 550,clip]{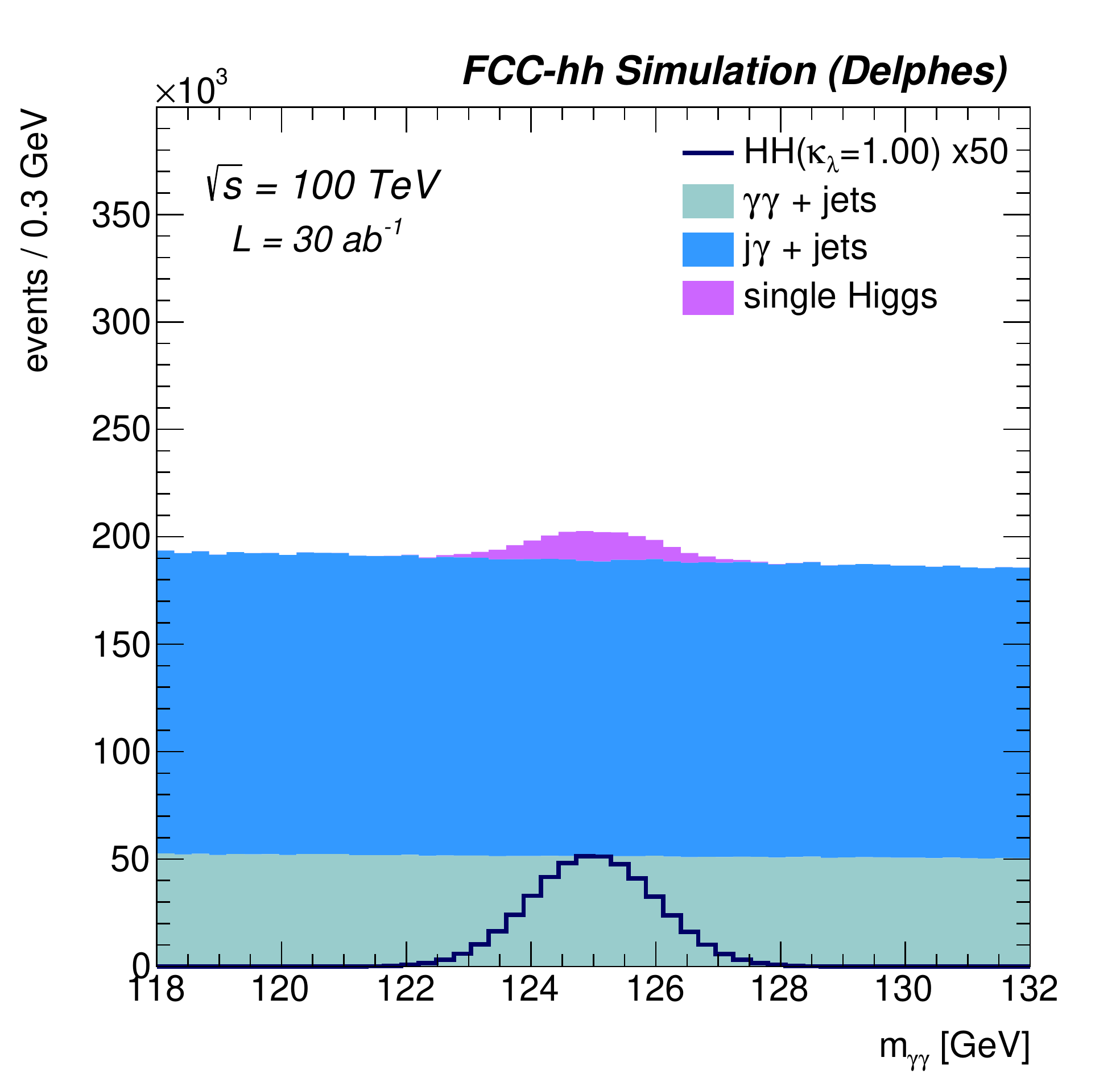}
  }
  \subfigure
  {\label{fig:hbb_m}
   \includegraphics[width = 0.31\linewidth,bb = 10 0 525 550,clip]{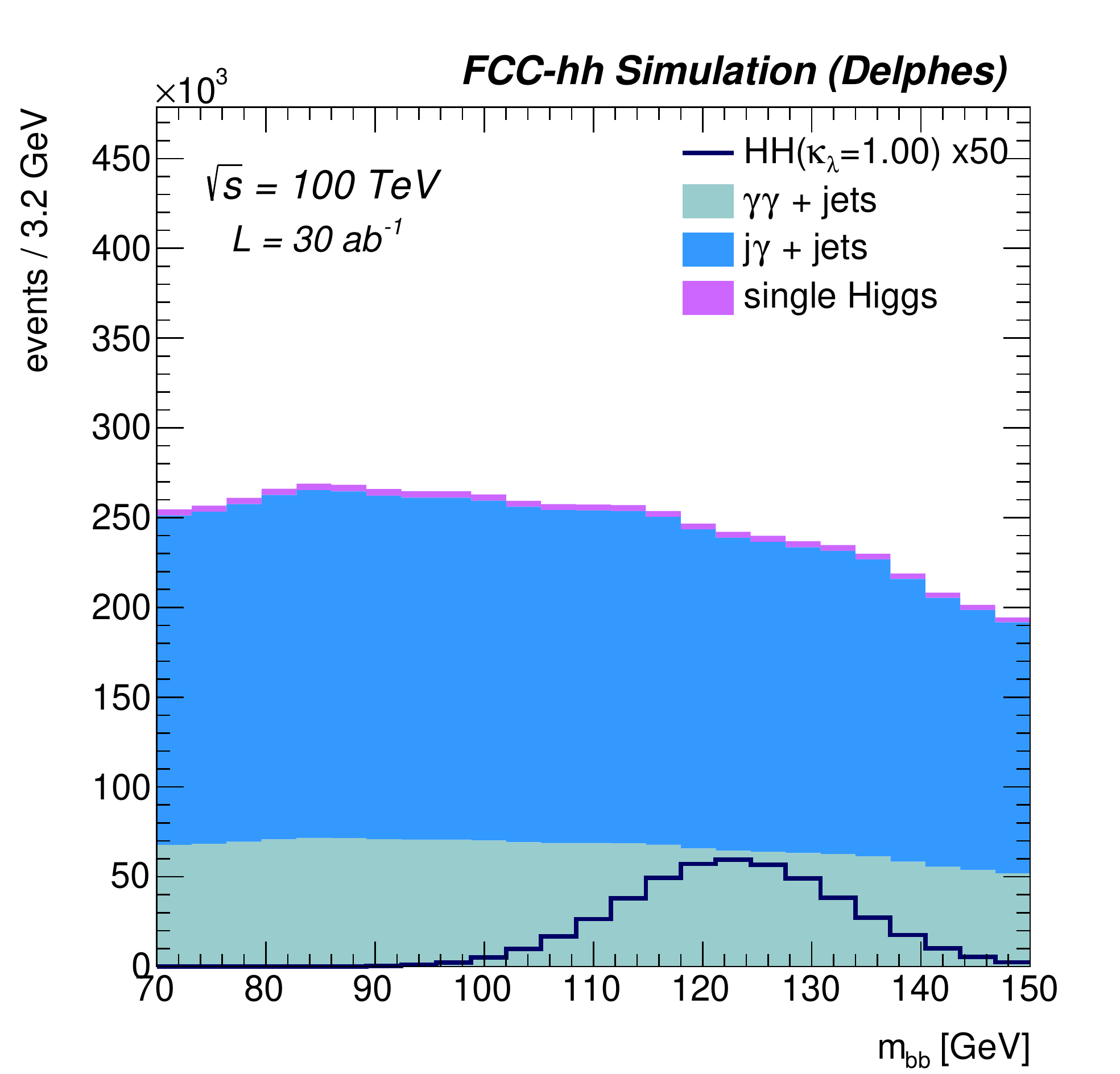}
  }
  \subfigure
  {\label{fig:hh_m}
   \includegraphics[width = 0.31\linewidth,bb = 10 0 525 550,clip]{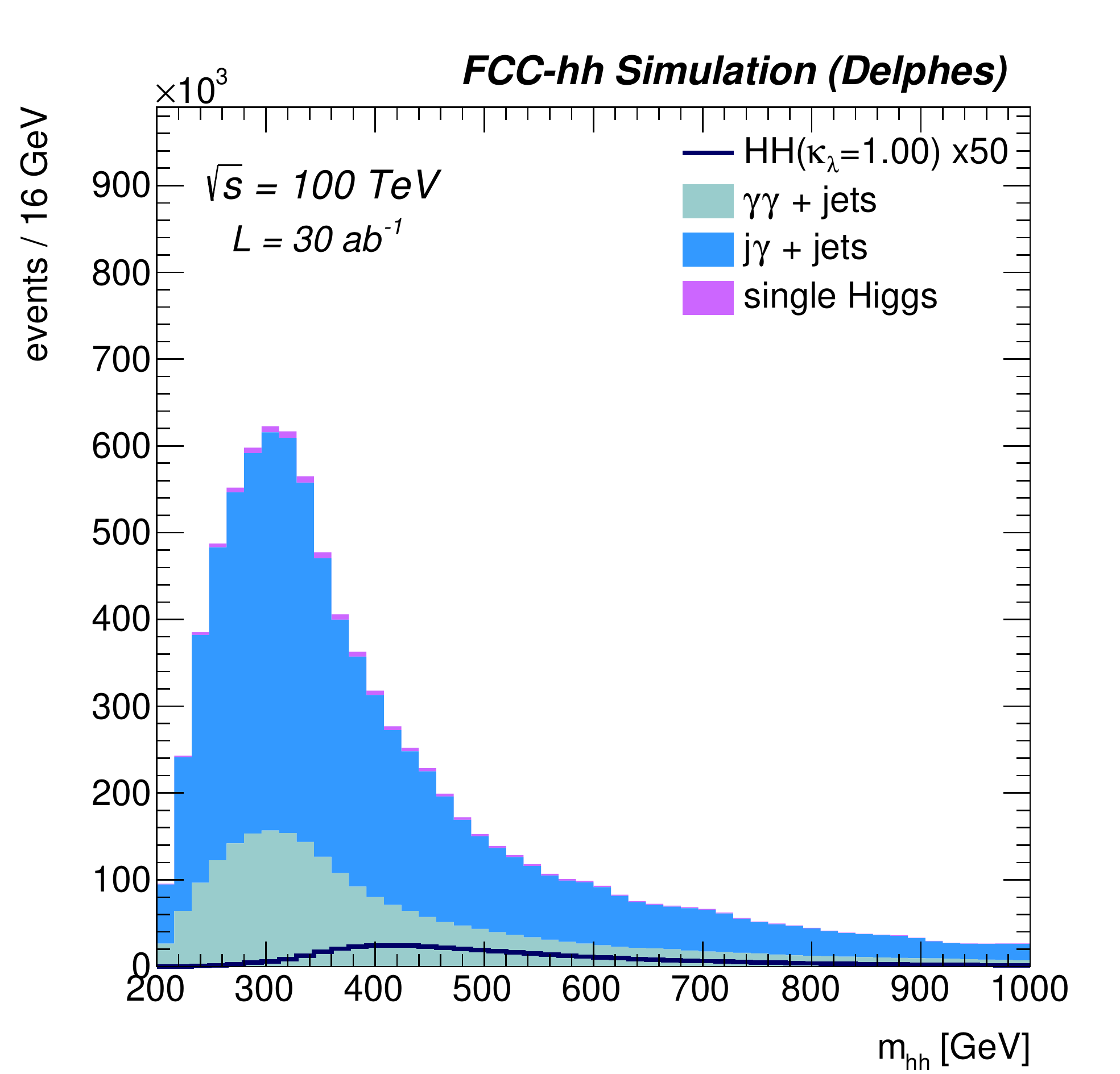}
  }
  \caption{Invariant mass spectra of the \haa~(a), \hbb~(b), HH~(c) candidates after applying the event pre-selection. The SM Higgs pair process is normalized to 50 times the expected yield with \intlumifcc.}
\end{figure}

In a future FCC-hh experiment, identification algorithms for photon and heavy-flavour will make use of the information of the invariant mass of the photon or jet candidate. Therefore we have to assume that the parameterised performance of the identification efficiency of such objects (in \delphes) already accounts for these variables. As a result,  the photon and jet mass are not used as input variables in the BDT discriminant. The \maa, \mbb, and \mhh\ observables, shown respectively in Figs.~\ref{fig:haa_m},~\ref{fig:hbb_m} and ~\ref{fig:hh_m} provide most of the discrimination against the background. 

The QCD (\aj\ and \aaj) and single-Higgs background processes possess different kinematic properties, and are therefore treated in separate classes. In QCD backgrounds, the final photons and jets tend to be softer and at higher rapidity. Conversely, the photon-pair candidates in single-Higgs processes often originate from a Higgs decay. As a result, while the \maa\ observable is highly discriminating against QCD, it is not against single-Higgs processes. In order to maximally exploit these kinematic differences we perform a separate training for each class of backgrounds, producing two multivariate discriminants: \bdth\ and  \bdtqcd. During the training, each background within each class is weighted according to the relative cross section.

\begin{figure}
\centering     
  \subfigure
  {\label{fig:sig_bdt2d}
   \includegraphics[width = 0.31\linewidth,,bb = 10 0 525 550,clip]{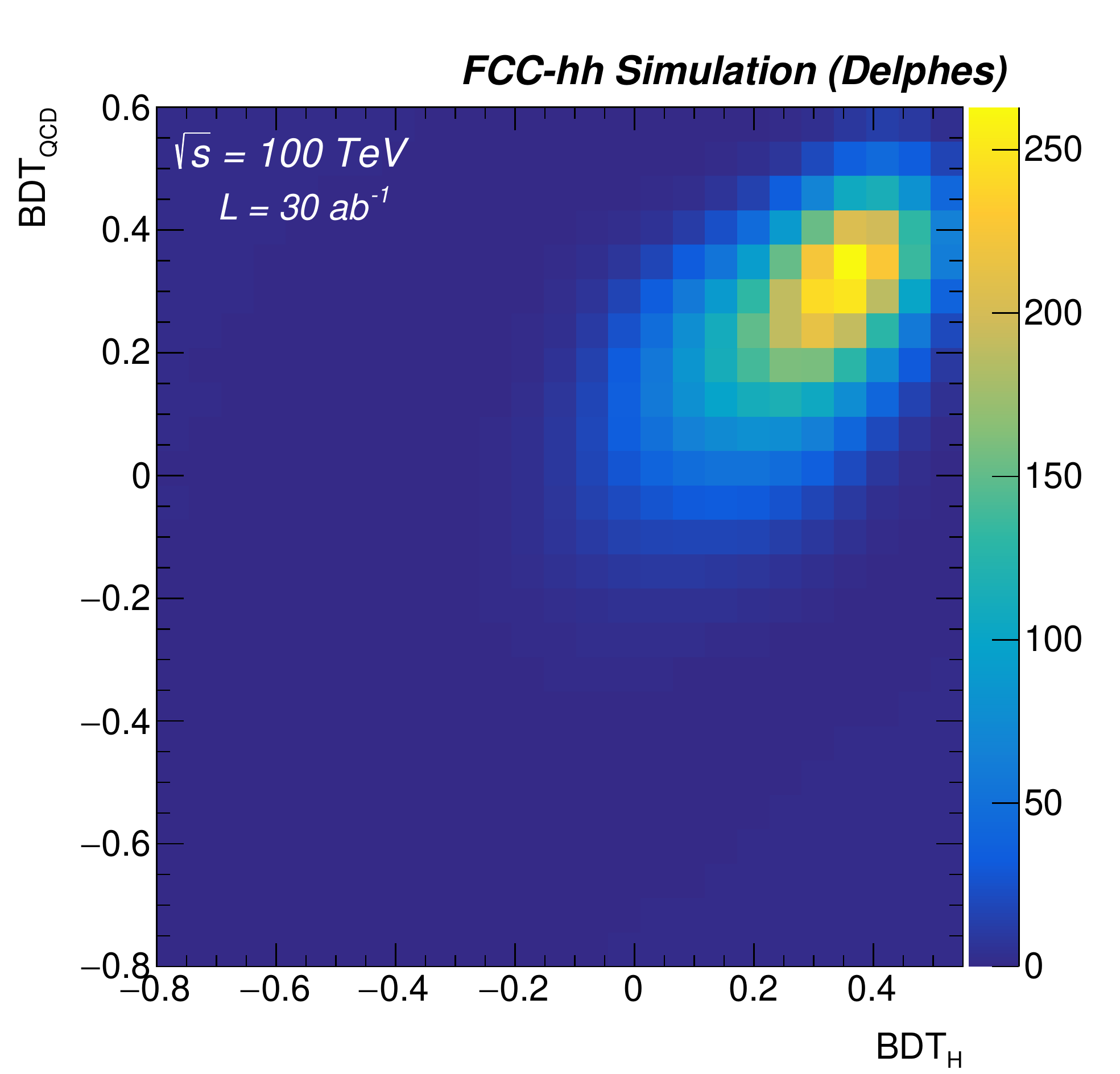}
  }
  \subfigure
  {\label{fig:jjaa_bdt2d}
   \includegraphics[width = 0.31\linewidth,,bb = 10 0 525 550,clip]{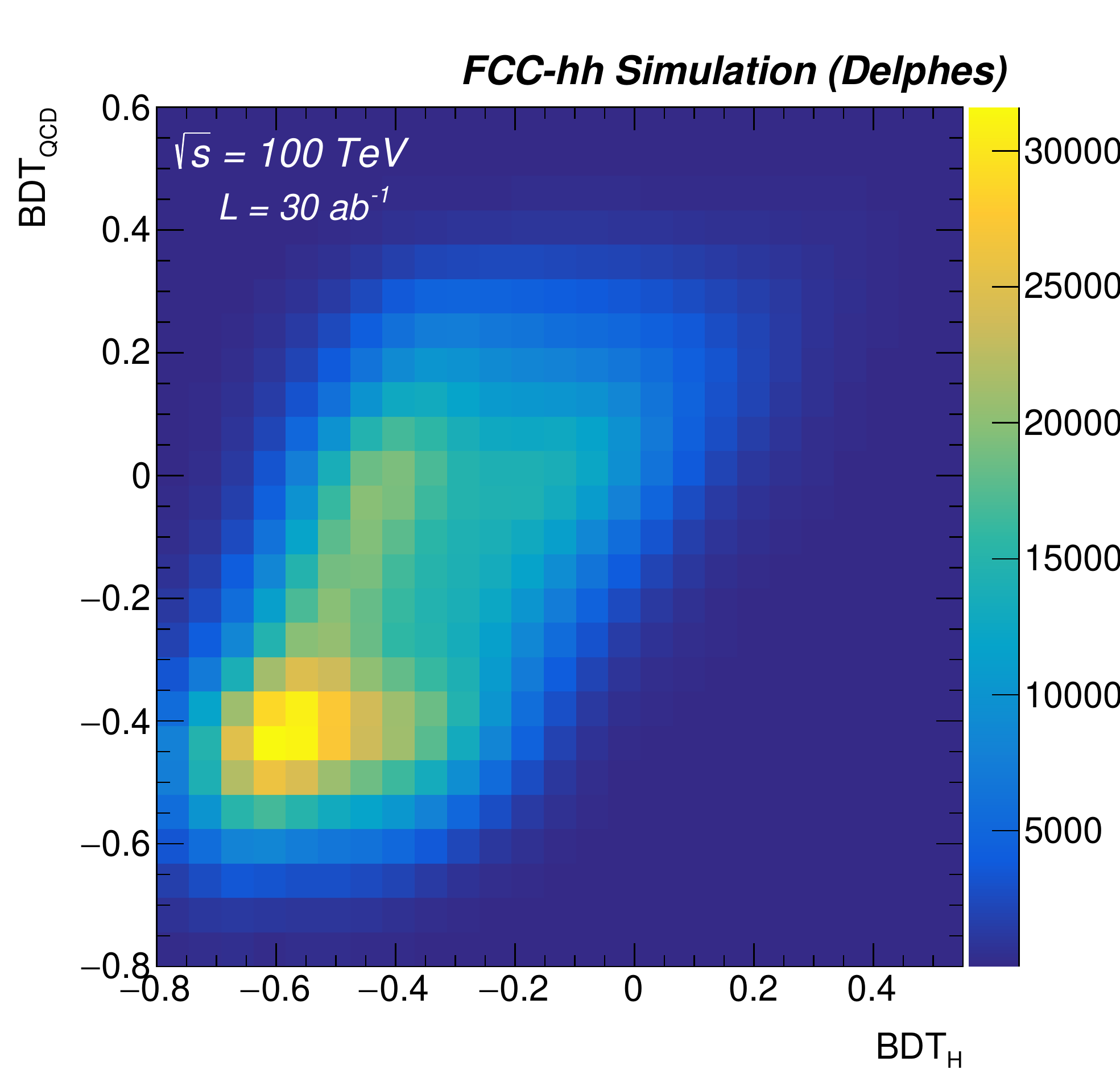}
  }
  \subfigure
  {\label{fig:singleh_bdt2d}
   \includegraphics[width = 0.31\linewidth,bb = 10 0 525 550,clip]{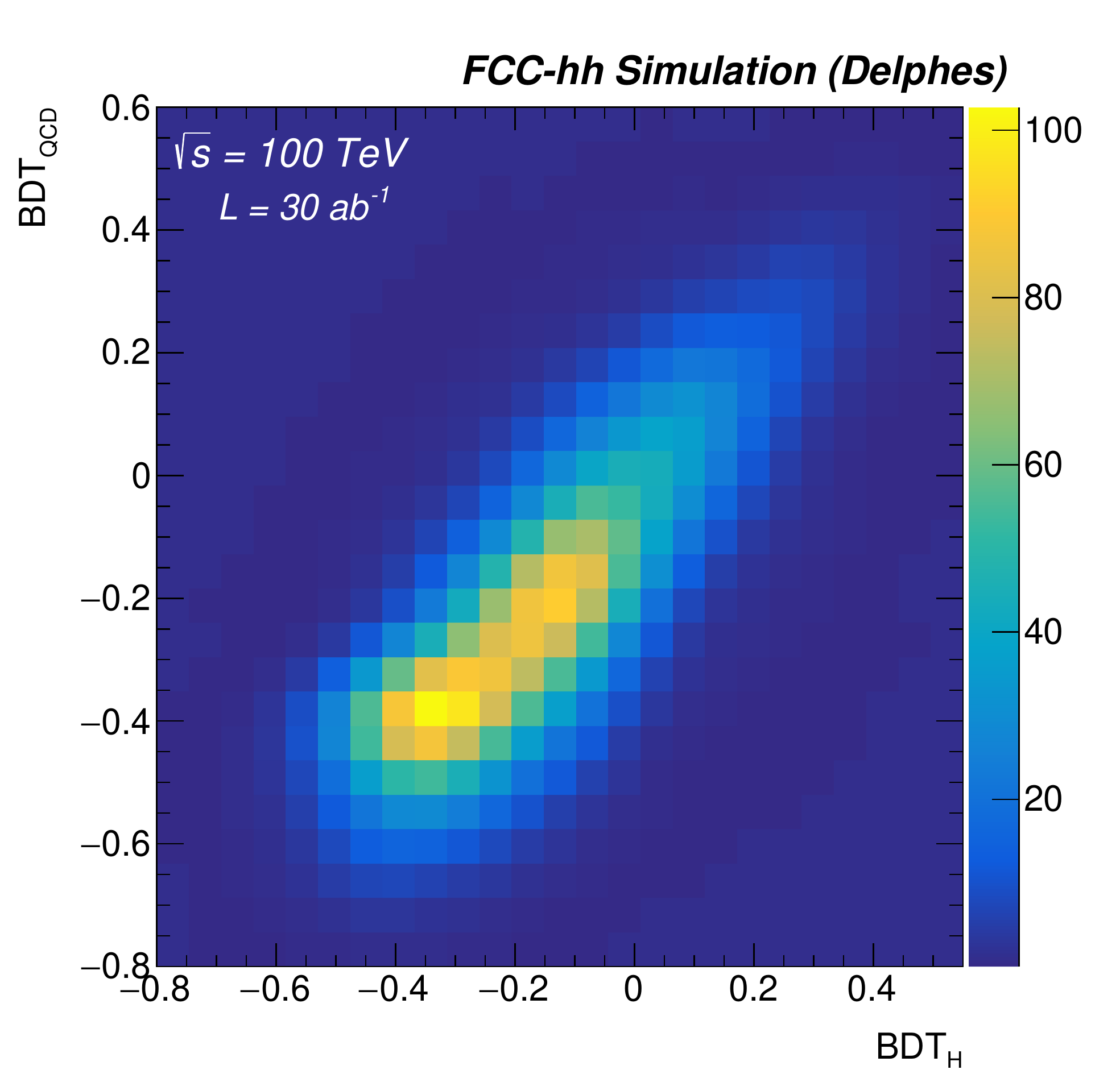}
  }
  \caption{Spectrum of SM signal (a), the QCD (b) and single Higgs (c) backgrounds in the (\bdth,\bdtqcd) plane.}
\end{figure}

The discrimination against single-Higgs backgrounds (\bdth) is largely driven by the \mbb\ variable, followed by \mhh\ and \ptSub{\bb}. Additional separation power, in particular against the \tth\ background, is provided by the number of reconstructed jets $N_j$ as well as \ptSup{\jmax} and \ptSup{\jmin}. On the other hand, the discrimination against QCD is driven by the \maa\ and \mbb\ variables, followed by \ptSub{\aa}, \ptSub{\bb}, \ptSup{\amax}, \ptSup{\bmax},  \mhh, \ptSup{\amin} and \ptSup{\bmin}.  

The output of the BDT discriminant is shown in the \bdtplane\ plane for the signal and the two background components in Figs.~\ref{fig:sig_bdt2d},~\ref{fig:jjaa_bdt2d} and ~\ref{fig:singleh_bdt2d}, respectively. As expected, the signal (background) enriched region clearly corresponds to large (small) \bdth\ and \bdtqcd\ values. We note that the multivariate discriminant correctly identifies the two main components (\ggh\ and \tth) within the single-Higgs background. The \ggh\ background, as opposed to \tth, is more ``signal-like'' and populates a region of high \bdth\ and \bdtqcd.

\subsubsection{Signal Extraction and results}
\begin{figure}[ht!]
\centering     
  \subfigure
  {\label{fig:bbaa_mu}
   \includegraphics[width = 0.47\linewidth,bb = 30 0 530 400,clip]{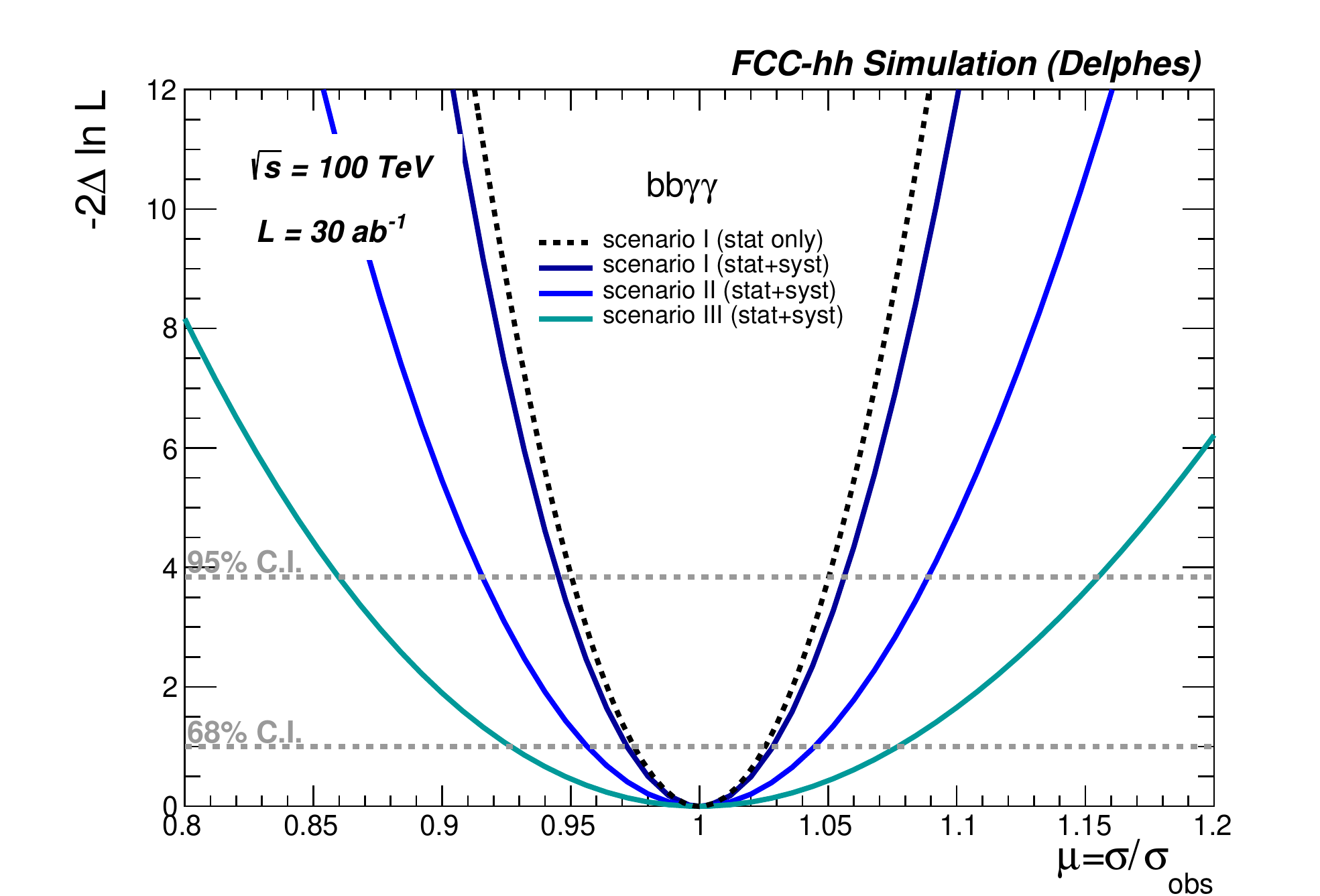}
  }
  \subfigure
  {\label{fig:bbaa_kl}
   \includegraphics[width = 0.47\linewidth,bb = 30 0 530 400,clip]{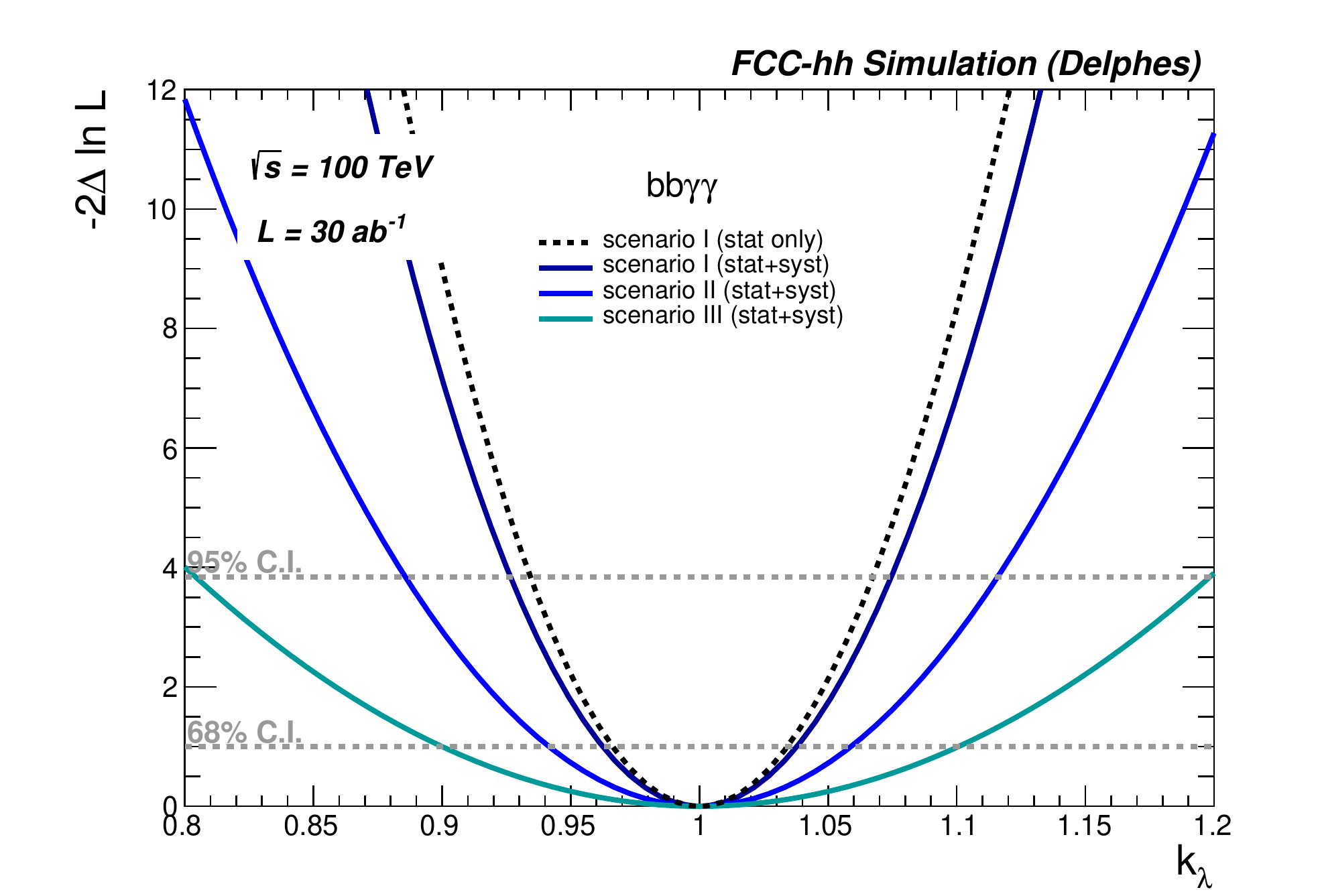}
}
  \caption{Expected negative log-Likelihood scan as a function the signal strenth \mufrac(a) and trilinear self-coupling modifier \klfrac(b) in the \bbaa\ channel. The various lines correspond to the different systematic uncertainty assumptions summarized in Table~\ref{tab:syst}. The black dashed line shows the likelihood profile when only the statistical uncertainty is included under scenario I.}
\end{figure}

The expected precision on the signal strength \mufrac\ and on the self-coupling modifier \klfrac\ are obtained from a 2-dimensional fit of the \bdtplane\ output, following the procedure described in Appendix~\ref{subsec:procedure}. The results are shown in Figs.~\ref{fig:bbaa_mu} and~\ref{fig:bbaa_kl}. The various lines correspond to the different scenarios described in Sections~\ref{subsec:reconstruction} and~\ref{subsec:systematics}. From Figure~\ref{fig:bbaa_kl} one can extract the symmetrized 68\% and 95\% confidence intervals for the various scenarios. The expected precision for  \bbaa\ is summarized in Table~\ref{tab:bbaa_res} for each of these assumptions. Depending on the assumed scenario, the Higgs self-coupling can be measured with a precision of 3.8-10\% at 68\% C.L using the \bbaa\ channel alone. We note that the achievable precision is largely dependent on the assumptions on the detector configuration and the systematic uncertainties. Such result needs to be compared against the precision of \dkl$=$3.4-7.4\%, obtained using statistical uncertainties alone. It is clear that the performance of the detector. In particular degrading the photon and jet energy resolution can have a substantial impact on the achievable precision.

%

\begin{table}
\centering
\begin{tabular}{cccc}

 @68\% CL  & scenario I & scenario II & scenario III \\
\hline 
\dmu\,\,\,\,  
    \begin{tabular}{ll}     stat only  \\    stat + syst \end{tabular}
&
    \begin{tabular}{@{}c@{}} 2.5 \\ 2.8 \end{tabular}
&
    \begin{tabular}{@{}c@{}} 3.6 \\ 4.4 \end{tabular}
&
    \begin{tabular}{@{}c@{}} 5.6 \\ 7.5 \end{tabular}
\\
\hline 

\dkl\,\,\,\, 
    \begin{tabular}{ll}     stat only  \\    stat + syst \end{tabular}
&
    \begin{tabular}{@{}c@{}} 3.4 \\ 3.8 \end{tabular}
&
    \begin{tabular}{@{}c@{}} 4.8 \\ 5.9 \end{tabular}
&
    \begin{tabular}{@{}c@{}} 7.4 \\ 10.0 \end{tabular}

\end{tabular}
\caption{\label{tab:bbaa_res}
Expected precision at 68\% CL on the di-Higgs production signal strength and Higgs self coupling using the \bbaa\ channel at the FCC-hh with \intlumifcc. The symmetrized value $\delta=(\delta^++\delta^-)/2$ is given in \%.}
\end{table}

\subsection{The \bbtata\ channel}
\label{subsec:bbtata}
The \bbtata\ channel is very attractive thanks to the large branching fraction (7.3\%) and the relatively clean final state. As opposed to the \bbaa\ channel, the \hhbbtata\ decay cannot be fully reconstructed due to the presence of $\tau$ neutrinos in the final state.
We consider mainly two channels here: the fully hadronic final state \bbtahtah, and the semi-leptonic one, \bbtahtal ($\ell=e,\mu$).

As spelled out in Section~\ref{subsec:mcb}, several processes act as background for the \bbtata\ final state. The largest background contributions are QCD and \ttbar. QCD is a background mainly for the \bbtahtah\ decay channel. However, the absence of prompt missing energy in QCD events makes this background reducible. We have verified that it can be suppressed entirely and therefore has been safely neglected here. Moreover, analyses using CMS data~\cite{Sirunyan:2017djm} show that QCD is overall a subdominant background at the LHC, and negligible in the signal region. As a result recent CMS Phase II projections neglect this background altogether~\cite{CMS-PAS-FTR-18-019}.
In order of decreasing magnitude, the largest backgrounds are \zjets\, single Higgs, ttV and ttVV, where V=W,Z.

\subsubsection{Event selection}

Events are required to contain at least two b-jets with \ptb~>~30~GeV and \etab~<~3.0. We require at least, and not exactly, two bjets in order not to suppress the \tthh\ signal contribution. For the \bbtahtal\ final state the presence is required of exactly one isolated ($\reliso~<~0.1$) lepton $\ell=e,\mu$ with \ptl~>~25~GeV and \etal~<~3.0 and at least one hadronically tagged $\tau$-jet with \pttah~>~45~GeV and \etatah~<~3.0. For the \bbtahtah\ final state, we require at least two hadronically tagged $\tau$-jet with \pttah~>~45~GeV and \etatah~<~3.0. The hadronic $\tau$ is identified according to the "Tight" criterion defined in Table~\ref{tab:detperf}. This ensures a highly efficient rejection of the QCD background (at the cost of a smaller $\tau_h$ efficiency) which strengthens the solidity of our assumption of neglecting the QCD background altogether. In what follows we refer to a $\tau$-candidate as the lepton $\ell=e,\mu$ or the $\tau$-jet. In particular the $\tau$ 4-momentum is defined as the sum of the 4-momenta of the visible $\tau$ decay products. In order to maximally exploit the kinematic differences between the signal and the dominant \ttbar\ background, we build a multivariate BDT discriminant using as an input the following kinematic properties:

\begin{itemize}
  \item The 3-vector components of the leading (\tamax) and subleading $\tau$-candidate (\tamin): transverse momentum (\ptSup{\tamax}, \ptSup{\tamin}), pseudo-rapidity (\etaSub{\tamax}, \etaSub{\tamin}), and azimutal angle (\phiSub{\tamax}, \phiSub{\tamin}).

  \item The 3-vector components of the leading (\bmax) and subleading b-jet (\bmin): transverse momentum (\ptSup{\bmax}, \ptSup{\bmin}), pseudo-rapidity (\etaSub{\bmax}, \etaSub{\bmin}), and azimutal angle (\phiSub{\bmax}, \phiSub{\bmin}).

  \item The 3-vector components of the leading (\jmax) and subleading additional reconstructed jets in the event (\jmin): transverse momentum (\ptSup{\jmax}, \ptSup{\jmin}), pseudo-rapidity (\etaSub{\jmax}, \etaSub{\jmin}), and azimutal angle (\phiSub{\jmax}, \phiSub{\jmin}). If no additional jets are found, dummy values are given to these variables. 

  \item The 4-vector components of the \htata\ candidate: transverse momentum (\ptSup{\tata}), pseudo-rapidity (\etaSub{\tata}), azimutal angle (\phiSub{\tata}) and invariant mass (\mtata).

  \item The 4-vector components of the \hbb\ candidate: transverse momentum (\ptSup{\bb}), pseudo-rapidity (\etaSub{\bb}), azimutal angle (\phiSub{\bb}) and invariant mass (\mbb).

  \item The 4-vector components of the Higgs pair candidate: transverse momentum (\ptSup{\hh}), pseudo-rapidity (\etaSub{\hh}), azimutal angle (\phiSub{\hh}) and invariant mass (\mhh).

  \item The transverse missing energy \metmag.

  \item The transverse mass of each $\tau$-candidate, computed as $\mt = \sqrt{2\ptSup{\tau} \metmag - \ptvecta \cdot \metvec}$.

  \item The event s-transverse mass \mttwo\ as defined in Refs~\cite{Lester:1999tx,Barr:2003rg}.
 
  \item The number of reconstructed b-jets $N_b$, the number of light jets $N_l$ and the total number of jets $N_j = N_b + N_l$.

\end{itemize}

\begin{figure}[t!]
  \centering     
    \subfigure[]
    {\label{fig:htahtah_m}
     \includegraphics[width = 0.31\linewidth,bb = 10 0 525 550,clip]{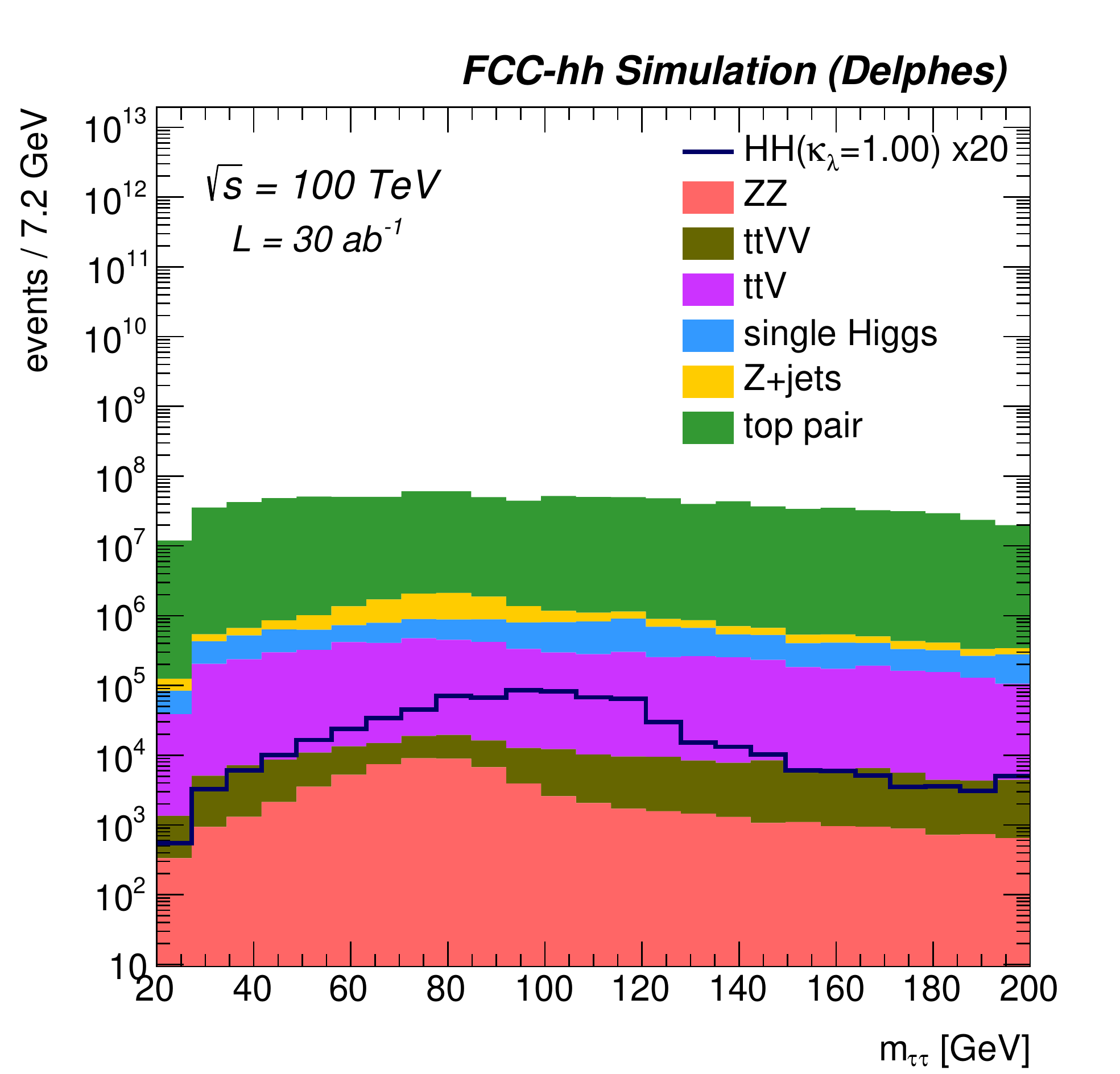}
    }
    \subfigure[]
    {\label{fig:tahtah_mt2}
     \includegraphics[width = 0.31\linewidth,bb = 10 0 525 550,clip]{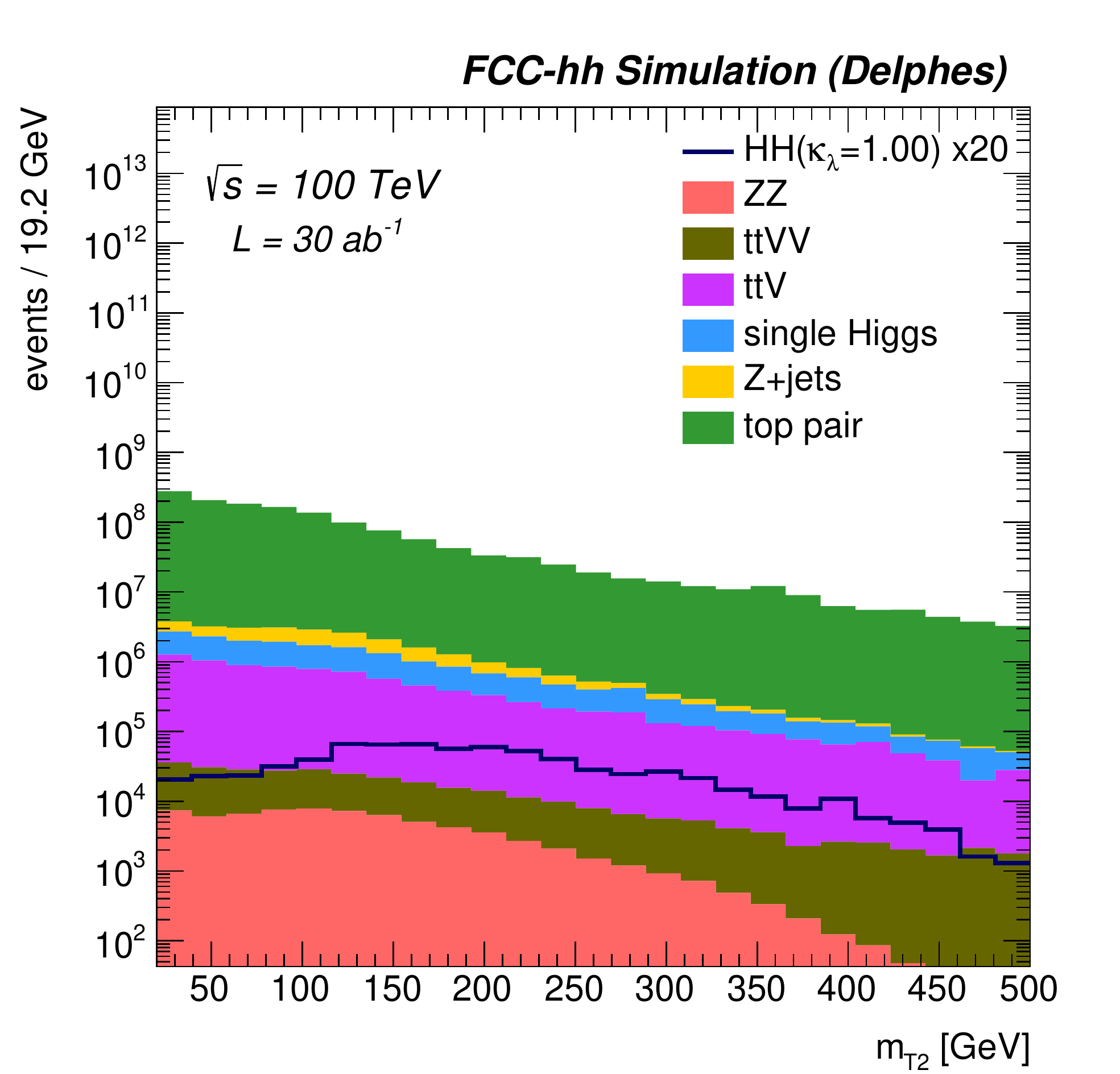}
    }
    \subfigure[]
    {\label{fig:bdttahtah}
     \includegraphics[width = 0.31\linewidth,bb = 10 0 525 550,clip]{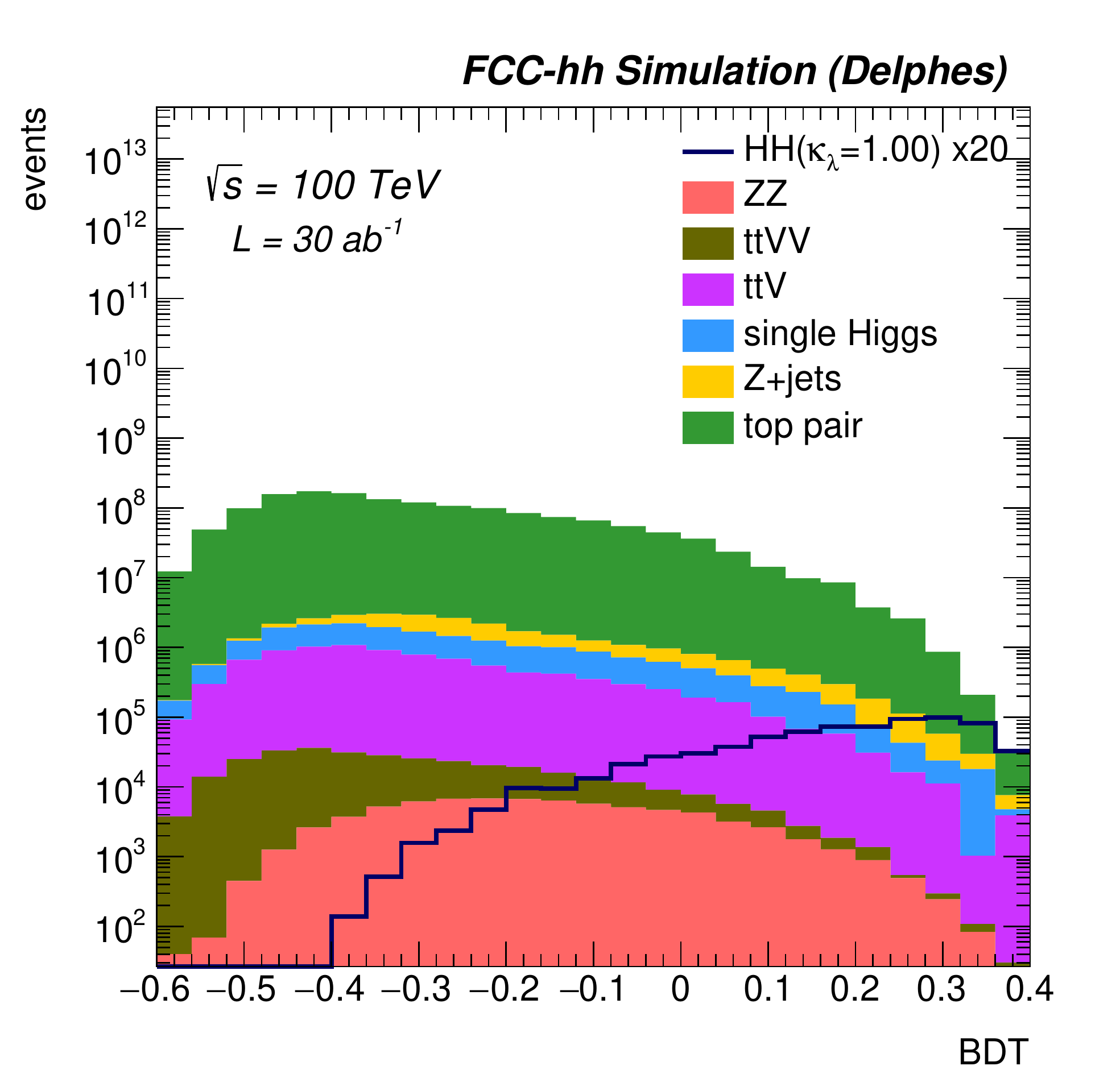}
    }
    \caption{Distributions in the \bbtahtah\ final state of the invariant mass of the \tata\ pair (left), \mttwo\ (center), and output of the BDT multi-variate discriminant (right).}
  \end{figure}

\begin{figure}[t!]
  \centering     
      \subfigure[]
      {\label{fig:htahtal_m}
       \includegraphics[width = 0.31\linewidth,bb = 10 0 525 550,clip]{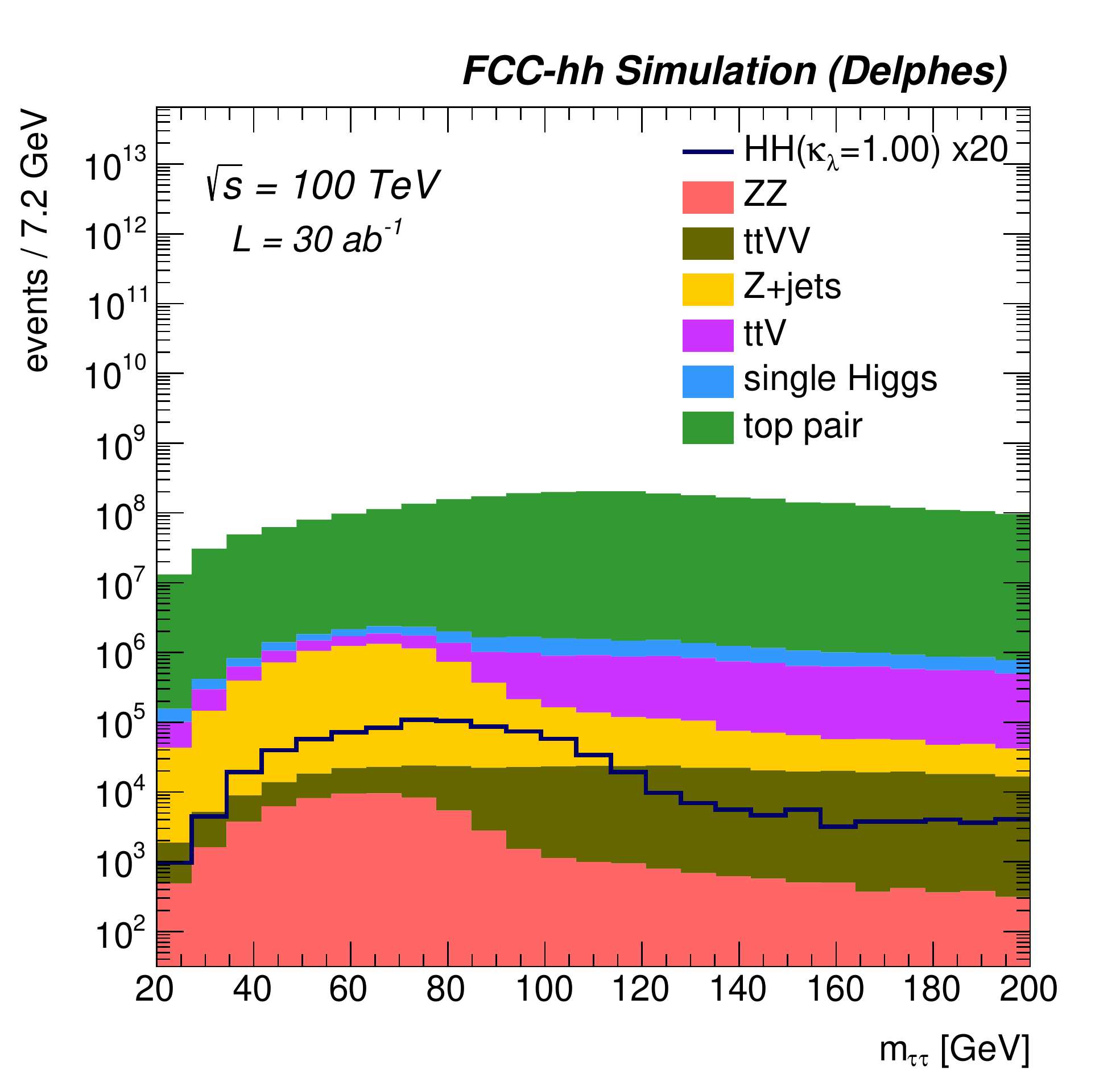}
      }
      \subfigure[]
      {\label{fig:tahtal_mt2}
       \includegraphics[width = 0.31\linewidth,bb = 10 0 525 550,clip]{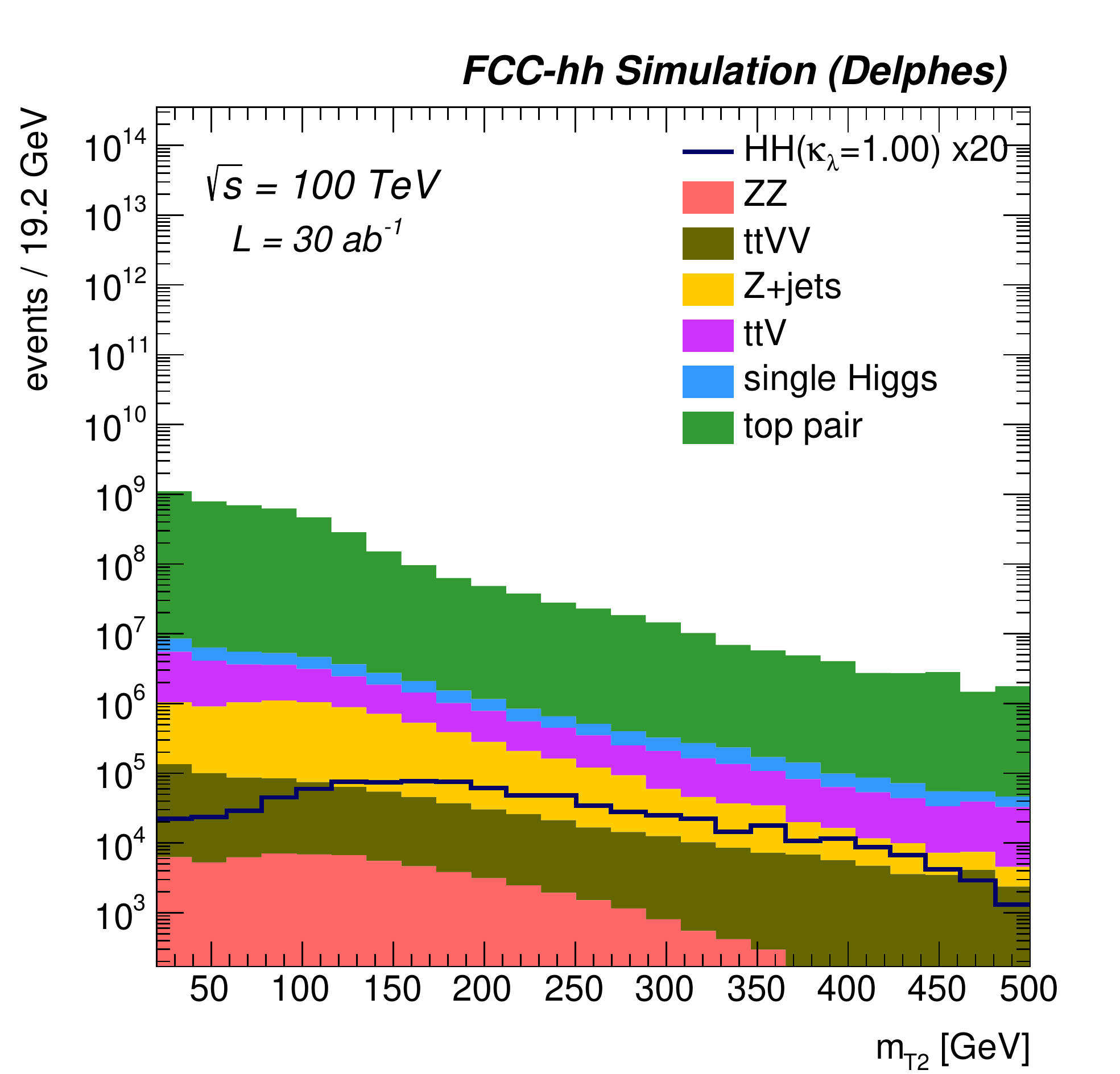}
      }
      \subfigure[]
      {\label{fig:bdttahtal}
       \includegraphics[width = 0.31\linewidth,bb = 10 0 525 550,clip]{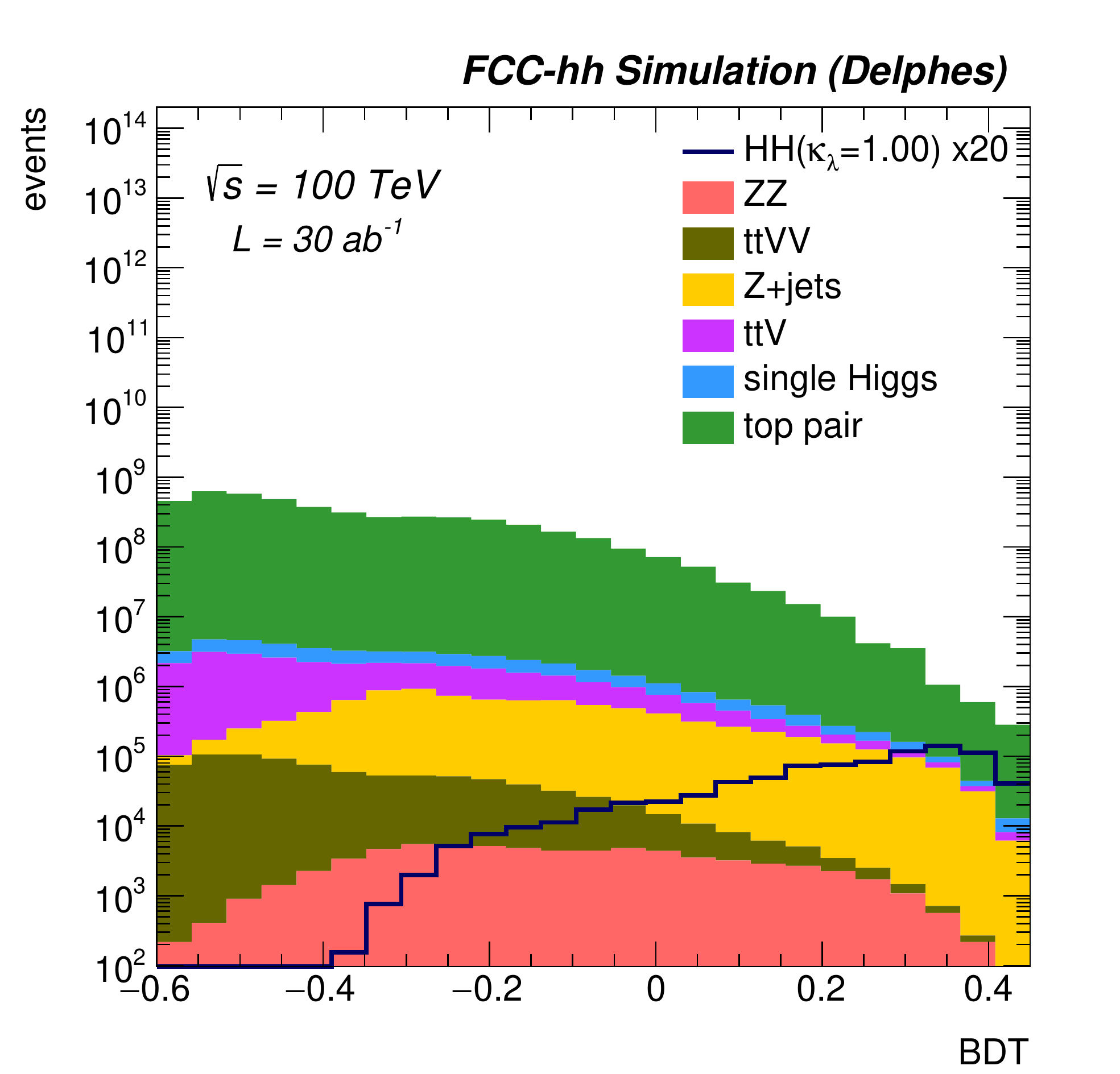}
      }
      \caption{Distributions in the \bbtahtal\ final state of the invariant mass of the \tata\ pair (left), \mttwo\ (center), and output of the BDT multi-variate discriminant (right).}
  \end{figure}

The \mtata\ and \mttwo\ observables are shown in Figs.~\ref{fig:htahtah_m},\ref{fig:htahtal_m} and~\ref{fig:tahtah_mt2},\ref{fig:tahtal_mt2} for the \bbtahtah\ and \bbtahtal\ final states respectively. These provide, together with the \mbb\ observable, the largest discrimination against the \ttbar\ background.  Additional discrimination against the \ttbar\ background is provided by \ptSup{\tata} and \ptSup{\bb} and the \metmag\, followed by \mhh, \etaSub{\tata}, \etaSub{\bb} and the \pt\ and $\eta$ of \tamin, \tamax, \bmin, \bmax, \jmin\ and \jmax, and finally $N_j$. The output of the BDT discriminant is shown in Figs.~\ref{fig:bdttahtah} and~\ref{fig:bdttahtal} for the \bbtahtah\ and \bbtahtal\ final states.

\subsubsection{Signal Extraction and results}

\begin{figure}[ht!]
\centering     
  \subfigure[]
  {\label{fig:bbtt_mu}
   \includegraphics[width = 0.47\linewidth,bb = 30 0 530 400,clip]{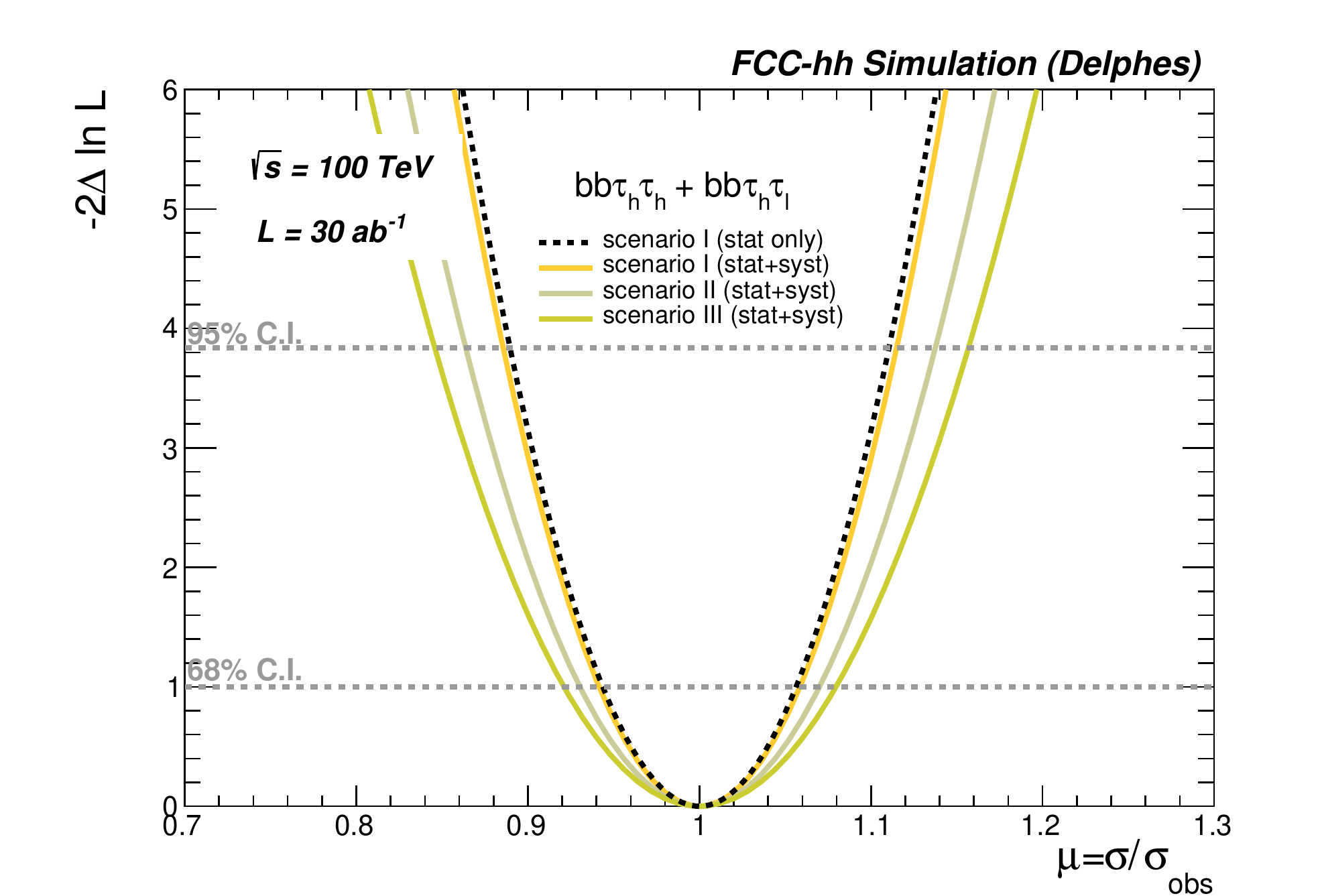}
  }
  \subfigure[]
  {\label{fig:bbtt_kl}
   \includegraphics[width = 0.47\linewidth,bb = 30 0 530 400,clip]{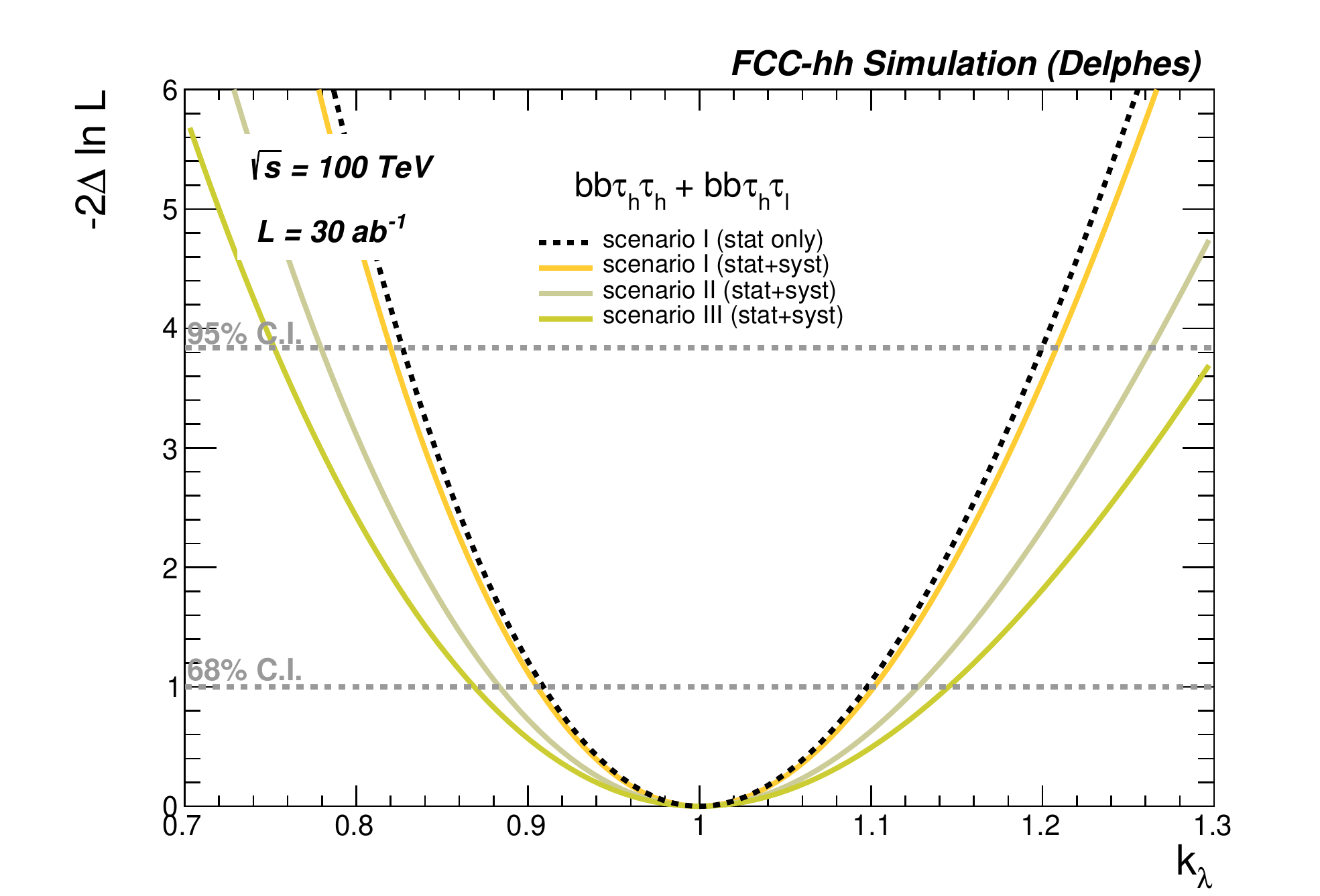}
  }
  \caption{Expected negative log-Likelihood scan as a function the signal strength \mufrac(a) and trilinear self-coupling modifier \klfrac(b) in the \bbtata\ channel (combination of the \bbtahtah\ and \bbtahtal\ channels). The various lines correspond to the different systematic uncertainties assumptions summarized in Table~\ref{tab:syst}. The black dashed line shows the likelihood profile when only the statistical uncertainty is included under scenario I.}
\end{figure}

The expected precision on the signal strength and the Higgs self-coupling are derived from a maximum likelihood fit on the BDT observable, according to the prescription described in Appendix~\ref{subsec:procedure}. The \bbtahtah\ and \bbtahtal\ channels are considered separately with their relative set of systematic uncertainties and then combined assuming a 100\% correlation on equal sources of uncertainties among the two channels. The combined expected precision on the \bbtata\ channel is shown in Figs.~\ref{fig:bbtt_mu} and~\ref{fig:bbtt_kl}.  The various lines correspond to the different scenarios described in Sections~\ref{subsec:reconstruction} and~\ref{subsec:systematics}. From Fig.~\ref{fig:bbtt_kl} one can extract the 68\% and 95\% confidence intervals for the various systematics assumptions. Depending on the assumed scenario, using the \bbtata\ channel, the Higgs pair signal strength and Higgs self-coupling can be measured respectively with a precision of $\dmu=5.8-7.9\%$ and $\dkl=9.8-13.8\%$ at 68\% C.L~\footnote{We note that the precision on the signal strength is higher for \bbtahtal\ compared to   \bbtahtah. The opposite is true of the self-coupling precision. This inversion can be explained by a smaller slope \slope\ (see Section~\ref{subsec:genstrat}) for the \bbtahtal\ channel, caused by different kinematic and acceptance requirements.}. Despite the large signal event rate in the \bbtata\ channel, the sensitivity is limited by the large background. Therefore, the \bbtata\ channel is statistically dominated at the FCC-hh with \intlumifcc, and the achievable precision is only moderately dependent on the assumptions on the systematic uncertainties. 
We note that Ref.~\cite{Banerjee:2018yxy} quotes a precision of $\dkl=13\%$ using the resolved semi-leptonic final state, which is consistent with the result presented here of $\dkl=14-18\%$. The same study shows that further precision can be obtained using the boosted topology, which has not been considered here.

\begin{table}
\centering
\begin{tabular}{cccc}

 @68\% CL   & scenario I & scenario II & scenario III \\
\hline 
\hline 
\dmu (\bbtahtah)\,\,\,\,  
\begin{tabular}{ll}     stat only  \\  stat + syst \end{tabular}
&
    \begin{tabular}{@{}c@{}} 8.7 \\ 9.0 \end{tabular}
&
    \begin{tabular}{@{}c@{}} 10.7 \\ 11.1 \end{tabular}
&
    \begin{tabular}{@{}c@{}} 10.9 \\ 11.6 \end{tabular}
\\
\hline 

\dkl (\bbtahtah)\,\,\,\, 
\begin{tabular}{ll}     stat only  \\  stat + syst \end{tabular}
&
    \begin{tabular}{@{}c@{}} 12.6 \\ 13.2 \end{tabular}
&
    \begin{tabular}{@{}c@{}} 16.0 \\ 16.7 \end{tabular}
&
    \begin{tabular}{@{}c@{}} 16.5 \\ 17.7 \end{tabular}

\\
\hline 
\hline 
\dmu (\bbtahtal)\,\,\,\,  
\begin{tabular}{ll}     stat only  \\  stat + syst \end{tabular}
&
    \begin{tabular}{@{}c@{}} 7.4 \\ 7.6 \end{tabular}
&
    \begin{tabular}{@{}c@{}} 8.2 \\ 8.8 \end{tabular}
&
    \begin{tabular}{@{}c@{}} 9.2 \\ 10.4 \end{tabular}
\\
\hline 

\dkl (\bbtahtal)\,\,\,\, 
\begin{tabular}{ll}     stat only  \\   stat + syst  \end{tabular}
&
    \begin{tabular}{@{}c@{}} 14.3 \\ 14.7 \end{tabular}
&
    \begin{tabular}{@{}c@{}} 16.3 \\ 17.5 \end{tabular}
&
    \begin{tabular}{@{}c@{}} 18.6 \\ 21.0 \end{tabular}

\\
\hline 
\hline 
\dmu (comb.)\,\,\,\,  
\begin{tabular}{ll}     stat only  \\   stat + syst  \end{tabular}
&
    \begin{tabular}{@{}c@{}} 5.6 \\ 5.8 \end{tabular}
&
    \begin{tabular}{@{}c@{}} 6.5 \\ 7.0 \end{tabular}
&
    \begin{tabular}{@{}c@{}} 7.0 \\ 7.9 \end{tabular}
\\
\hline 

\dkl (comb.)\,\,\,\, 
\begin{tabular}{ll}     stat only  \\   stat + syst \end{tabular}
&
    \begin{tabular}{@{}c@{}} 9.4 \\ 9.8 \end{tabular}
&
    \begin{tabular}{@{}c@{}} 11.4 \\ 12.2 \end{tabular}
&
    \begin{tabular}{@{}c@{}} 12.3 \\ 13.8 \end{tabular}

\\

\end{tabular}
\caption{\label{tab:bbtata_res}
Expected precision at 68\% CL on the di-Higgs production cross-section and Higgs self coupling using the \bbtata\ channel at the FCC-hh with \intlumifcc. The symmetrized value $\delta=(\delta^++\delta^-)/2$ is given in \%.}

\end{table}

\subsection{The \bbbb\ channel}
\label{subsec:bbbb}
The \hhbbbb\ decay mode has the largest branching fraction among all possible Higgs-pair decays. Despite the presence of soft neutrinos from semi-leptonic b decays (that may impact negatively the reconstructed hadronic Higgs-mass resolution), the Higgs decays into b-jets can be fully reconstructed. However, due do the fully hadronic nature of this decay mode, this channel suffers from the presence of very large QCD backgrounds and hence features a relatively small \sob. Moreover, a combinatorial ambiguity affects the possibility to correctly associate the four b-jets to the two parent Higgs candidates.

We consider mainly the case where the Higgs candidates are only moderately boosted, leading to four fully resolved b-jets. The boosted analysis, where the Higgs candidates are sufficiently boosted to decay into a single large radius jet~\cite{Behr:2015oqq,deLima:2014dta}, provides less sensitivity to the self-coupling measurement and was discussed in previous studies\cite{Banerjee:2018yxy, L.Borgonovi:2642471}. The main backgrounds to this final state are QCD and \ttbar, followed by \zbb, single-Higgs production and ZZ.

\subsubsection{Event selection}
In order to fulfill our initial assumption of fully efficient online triggers, the event selection starts by requiring the presence of at least four b-jets with \ptb~>~30~GeV and \etab~<~4.0. The b-jets are identified with the "Medium" working point defined in Table~\ref{tab:detperf}. The Higgs candidates are reconstructed as the pairing of b-jet pairs that minimizes the difference between the invariant masses of the two b-jet pairs. The Higgs candidate with the largest (smallest) \pt\ is named \hone (\htwo).

The following variables are then used as input to a multivariate BDT discriminant to ensure an optimal discrimination versus the dominant QCD background:
\begin{itemize}

\item The 3-vector components of the four leading b-jets in the event (\bone, \btwo, \bthree, \bfour): transverse momenta (\ptSup{\bi}), pseudo-rapidities (\etaSub{\bi}), and azimutal angles (\phiSub{\bi}, \onetofour).

 \item The 3-vector components of the leading (\jmax) and subleading additional reconstructed jets in the event (\jmin): transverse momentum (\ptSup{\jmax}, \ptSup{\jmin}), pseudo-rapidity (\etaSub{\jmax}, \etaSub{\jmin}), and azimutal angle (\phiSub{\jmax}, \phiSub{\jmin}). If no additional jets are found, dummy values are given to these variables. 

\item The 4-vector components of the leading (\hone) and subleading (\htwo) Higgs candidates: transverse momentum (\ptSup{\hone}, \ptSup{\htwo}), pseudo-rapidity (\etaSub{\hone}, \etaSub{\htwo}), azimutal angle (\phiSub{\hone}, \phiSub{\htwo}) and invariant mass (\mSub{\hone}, \mSub{\htwo}).

\item The 4-vector components of the Higgs-pair candidate: transverse momentum (\ptSup{\hh}), pseudo-rapidity (\etaSub{\hh}), azimutal angle (\phiSub{\hh}) and invariant mass (\mhh).

\end{itemize}

\begin{figure}[t!]
\centering     
  \subfigure[]
  {\label{fig:hbb1_m}
   \includegraphics[width = 0.31\linewidth,bb = 10 0 525 550,clip]{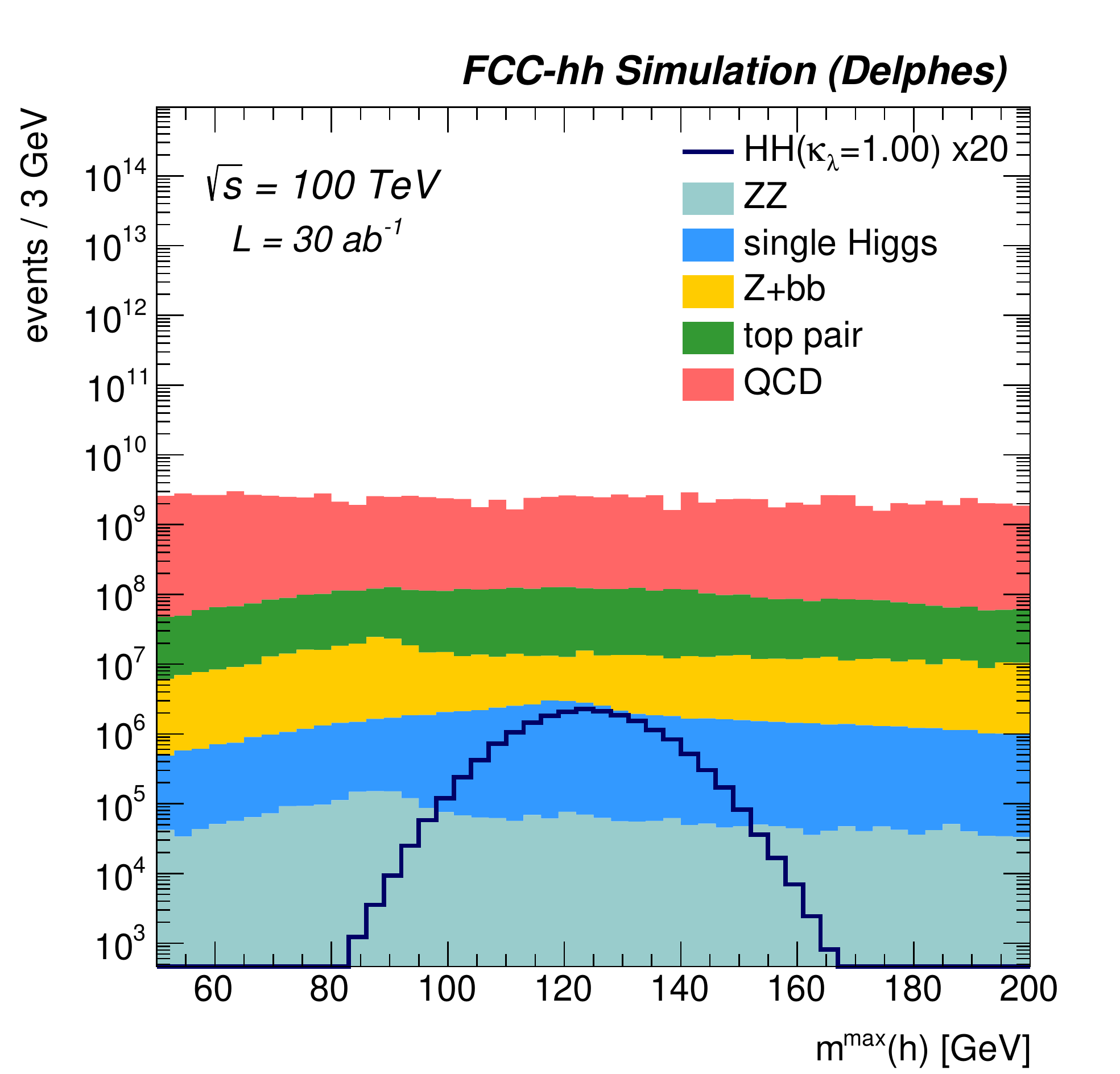}
  }
  \subfigure[]
  {\label{fig:bbbb_m}
   \includegraphics[width = 0.31\linewidth,bb = 10 0 525 550,clip]{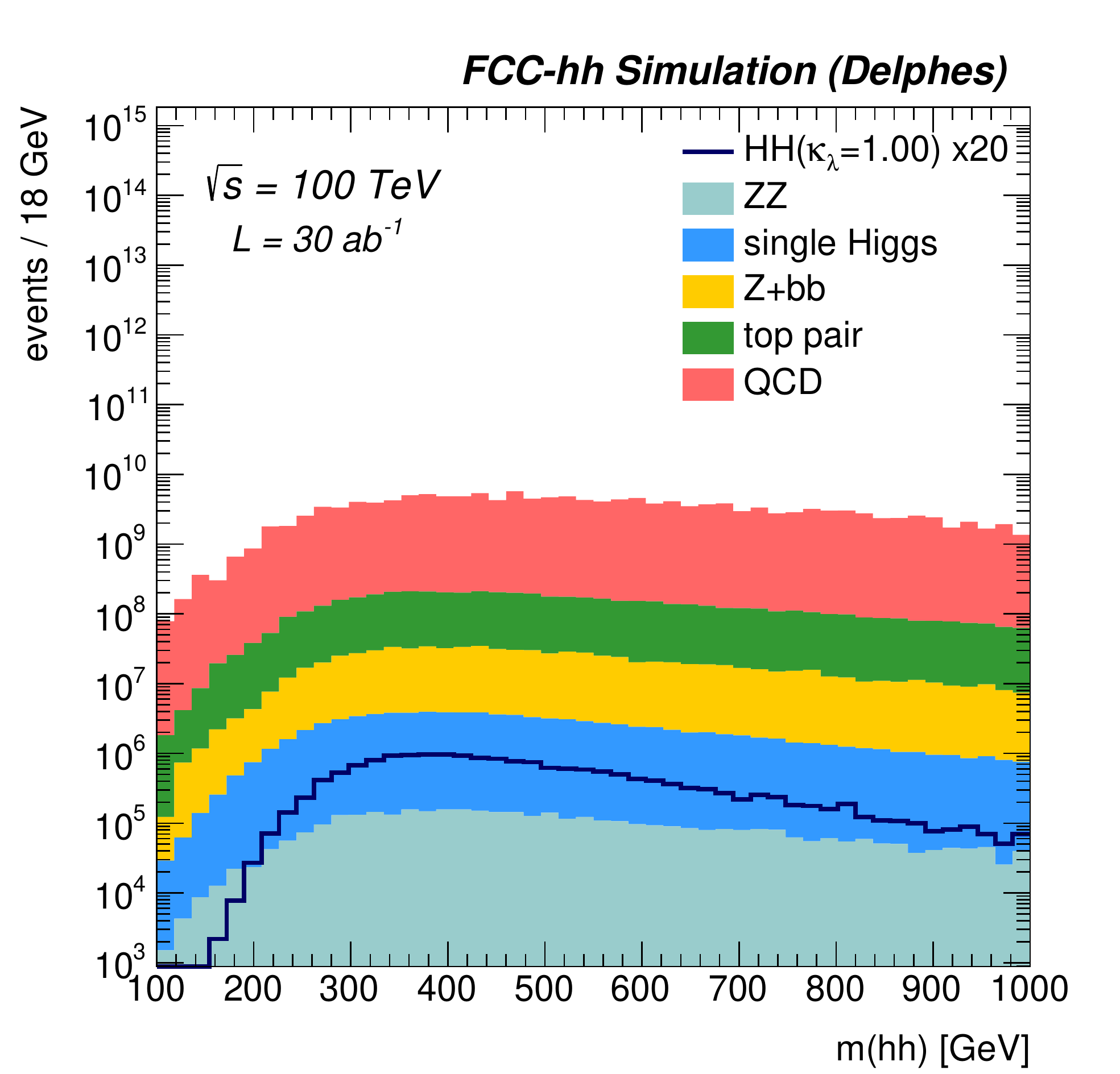}
  }
  \subfigure[]
  {\label{fig:btdbb}
   \includegraphics[width = 0.31\linewidth,bb = 10 0 525 550,clip]{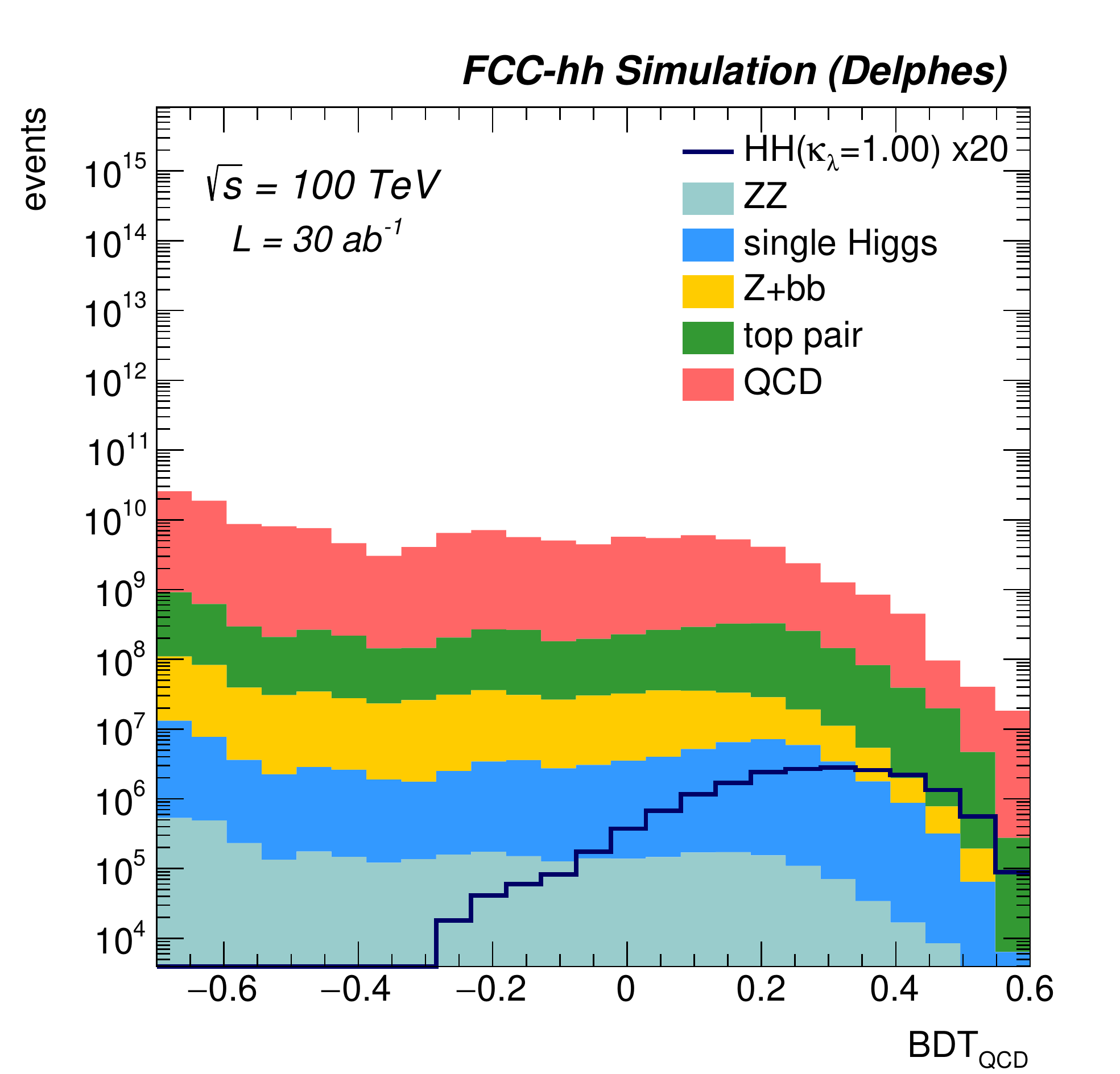}
  }
  \caption{Distributions in the \bbbb\ final state of the highest \pt\ reconstructed Higgs invariant mass (left), the Higgs pair invariant mass (center), and the output of the BDT multi-variate discriminant (right).}
\end{figure}

The \mSub{\hone} and \mhh\ observables are shown respectively in Figs.~\ref{fig:hbb1_m} and~\ref{fig:bbbb_m}. The \mSub{\hone} distribution shows that the procedure described above correctly associates the b-jet pairs to the parent Higgs particle for the signal, and to the parent Z particle for the \zbb\ and ZZ backgrounds. Thanks to their resonant nature, the \mSub{\hone} and \mSub{\htwo} distributions provide the largest discrimination against the QCD background. A substantial discrimination against the QCD background is provided, in decreasing order of importance, by \ptSup{\btwo}, \ptSup{\bfour}, \ptSup{\bthree} and \ptSup{\bone}, followed by \mhh, \ptSup{\hh}, \ptSup{\jmax}, \ptSup{\jmin}, and the \etaSub{\bi}. The output of the BDT discriminant is shown in Fig.~\ref{fig:btdbb} for the signal and various background contributions.

\subsubsection{Results}

\begin{figure}[ht!]
\centering     
  \subfigure[]
  {\label{fig:bbbb_mu}
   \includegraphics[width = 0.47\linewidth,bb = 30 0 530 400,clip]{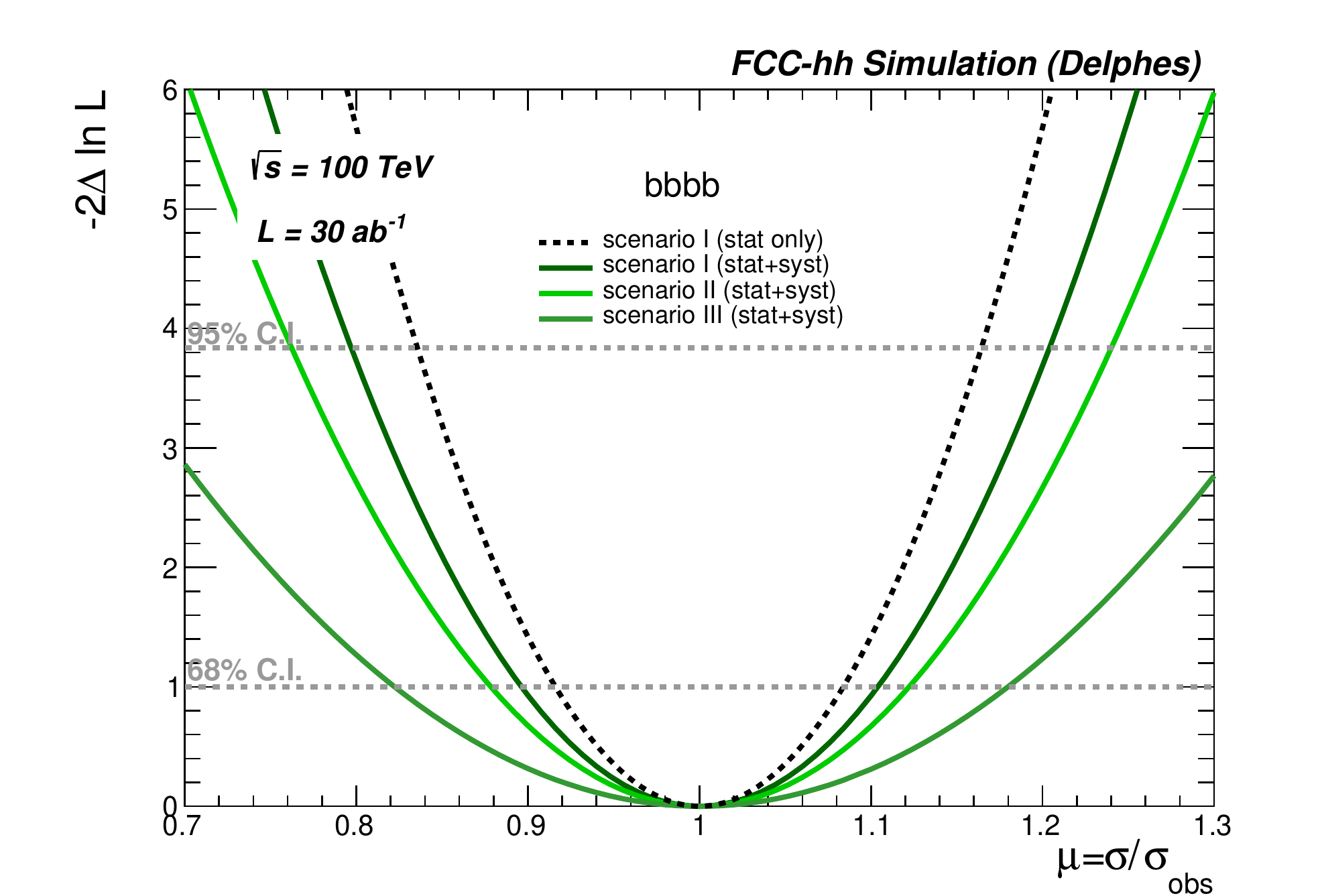}
  }
  \subfigure[]
  {\label{fig:bbbb_kl}
   \includegraphics[width = 0.47\linewidth,bb = 30 0 530 400,clip]{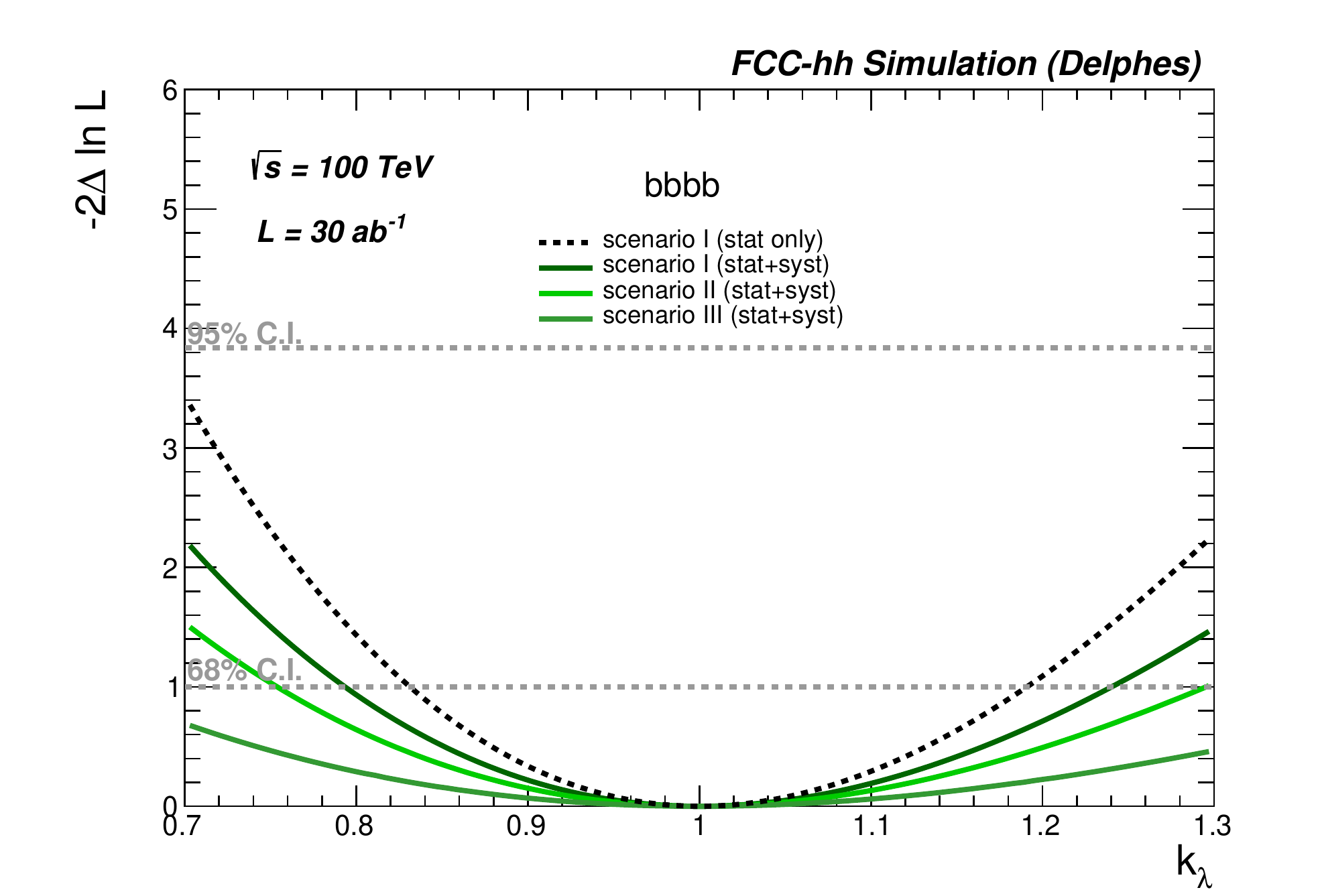}
  }
  \caption{Expected Negative log-Likelihood scan as a function the signal strength \mufrac(a) and trilear self-coupling modifier \klfrac(b) in the \bbbb\ channel. The various lines correspond to the different systematic uncertainties assumptions summarized in Table~\ref{tab:syst}. The black dashed line shows the likelihood profile when onlythe statistical uncertainty is included under scenario I.}
\end{figure}

The expected precision on the signal strength and the Higgs self-coupling are derived from a 1D maximum likelihood fit on the BDT discriminant, according to the prescription described in Section~\ref{subsec:procedure}. The combined expected precision on the \bbbb\ channel is shown in Figs.~\ref{fig:bbbb_mu} and~\ref{fig:bbbb_kl}. The coloured lines correspond to the different systematic uncertainties assumptions summarized in Table~\ref{tab:syst}. The 68\% and 95\% confidence intervals on \dmu\ and \dkl\ for the various systematics assumptions can be extracted from Figs.~\ref{fig:bbbb_mu} and~\ref{fig:bbbb_kl} and the results are summarized in Table~\ref{tab:bbbb_res}. Depending on the assumed scenario, using the \bbbb\ channel, the Higgs pair signal strenth and Higgs self-coupling can be measured respectively with a precision of $\dmu=8-18\%$ and $\dkl=18-32\%$ at 68\% C.L. As for the \bbtata\ case, due to the huge QCD background this channel is statistically limited. We note that, despite the large statistical uncertainty, the systematic uncertainties play a large role in this channel. This is the result of the significant contamination in the signal region from single-Higgs background events. Since this background is assumed to be estimated from Monte Carlo, its uncertainty, even though at the percent level, is reflected in a larger signal systematics.

\begin{table}
\centering
\begin{tabular}{cccc}

 @68\% CL   & scenario I & scenario II & scenario III \\
\hline 
\dmu\,\,\,\,  
\begin{tabular}{ll}     stat only  \\    stat + syst \end{tabular}
&
    \begin{tabular}{@{}c@{}} 8.4 \\ 10.4 \end{tabular}
&
    \begin{tabular}{@{}c@{}} 9.1 \\ 12.2 \end{tabular}
&
    \begin{tabular}{@{}c@{}} 10.8 \\ 17.9 \end{tabular}
\\
\hline 

\dkl\,\,\,\, 
\begin{tabular}{ll}     stat only  \\     stat + syst \end{tabular}
&
    \begin{tabular}{@{}c@{}} 18.0 \\ 22.3 \end{tabular}
&
    \begin{tabular}{@{}c@{}} 20.0 \\ 27.1 \end{tabular}
&
    \begin{tabular}{@{}c@{}} 24.2 \\ 32.0 \end{tabular}

\end{tabular}
\caption{\label{tab:bbbb_res} Expected precision at 68\% CL on the di-Higgs production cross-section and Higgs self coupling using the \bbbb\ channel at the FCC-hh with \intlumifcc. The symmetrized value $\delta=(\delta^++\delta^-)/2$ is given in \%.}
\end{table}

\subsection{Combined precision}
\label{subsec:comb}

\begin{figure}[ht!]
\centering     
  \subfigure
  {
   \includegraphics[width = 0.80\linewidth]{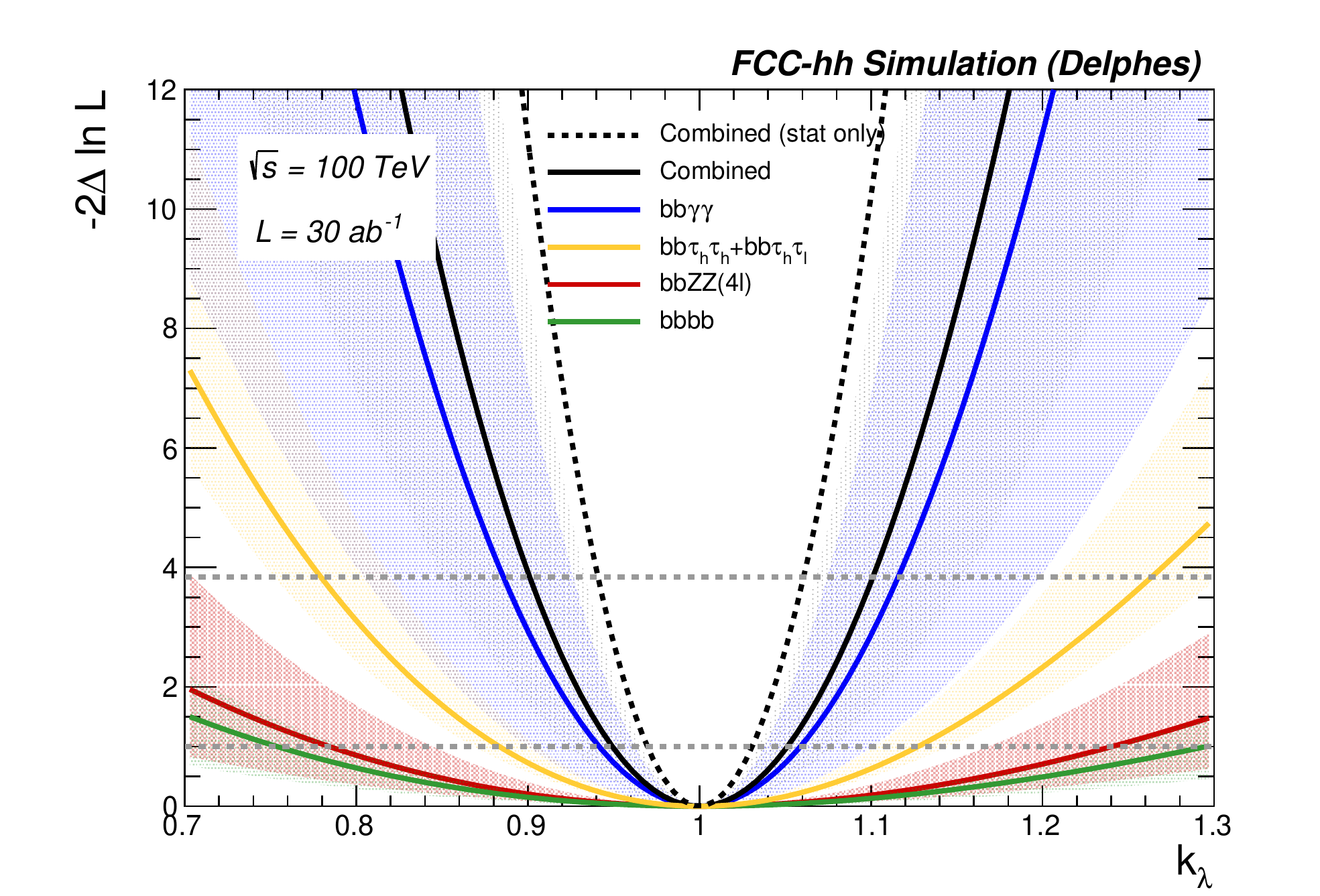}
  }
  \caption{Expected negative log-Likelihood scan as a function of the trilinear self-coupling modifier \klfrac\ in all channels, and their combination. The solid line corresponds to the scenario II for systematic uncertainties. The band boundaries represent respectively scenario I and III. The dashed line represents the sensitivity obtained including statistical uncertainties only, under the assumptions of scenario I.}
  \label{fig:comb_kl}
\end{figure}

\begin{figure}[ht!]
\centering     
  \subfigure
  {
   \includegraphics[width = 0.80\linewidth]{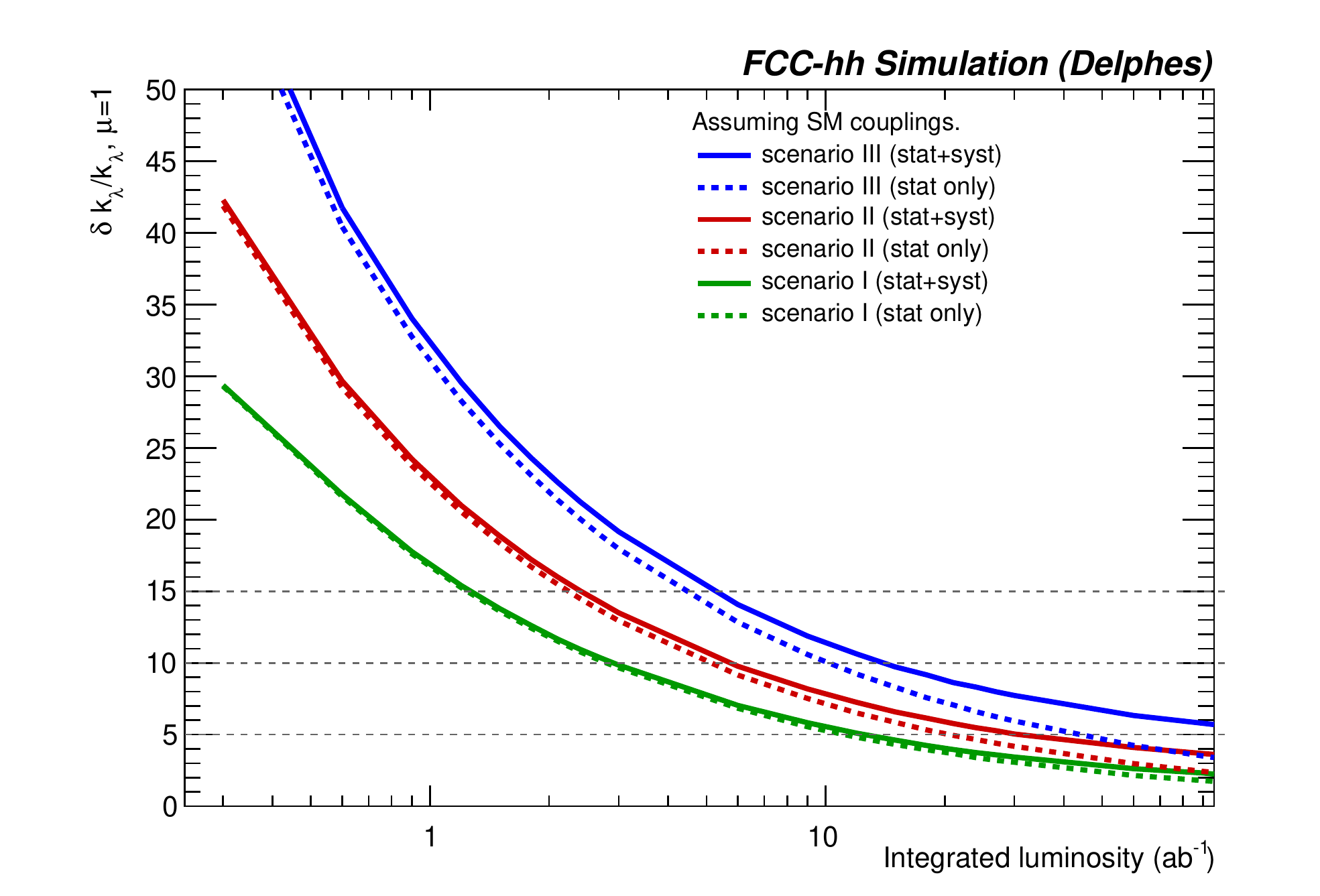}
  }
  \caption{Expected precision on the Higgs self-coupling as a function of the integrated luminosity.}
  \label{fig:comb_lumi}
\end{figure}

When combining results from the various channels, the systematic uncertainties from the various sources affecting those processes that we assume to be estimated from Monte Carlo simulations (HH, single Higgs, and ZZ) are accounted for as follows.\\
Lepton (e/$\mu$, $\tau$) uncertainties are correlated across all process and across the \bbtahtah\ and \bbtahtal\ channels. In the \bbaa\ channel, the photon uncertainty for the single and double Higgs processes are correlated. The luminosity uncertainty is correlated across all these processes and all channels. The same applies, for each process independently, to the overall normalisation uncertainty. The uncertainties on lepton ID, luminosity and normalisation are assumed to affect only the overall normalization of signal and background shapes and to not introduce a significant deformation of their shapes. For the b-jets ID, we take into account both the shape and normalisation uncertainty due to the b-jets systematic uncertainties. This is achieved by shifting the efficiency for each jet by the (\pt-dependent) values reported in Table~\ref{tab:detperf}, and re-computing the resulting BDT distributions. As discussed in section \ref{subsec:systematics}, this uncertainty is correlated across different channels when it affects the same process.
Overall normalisation uncertainties are cancelled out when a background is estimated from a control region. Moreover, we expect all control regions to be well-populated at the FCC-hh. For these reasons, we assume the systematic uncertainties affecting those backgrounds that are most likely to be estimated from well populated control regions or side bands, such as QCD, Zbb, photon(s)+jets, and \ttbar,  to be negligible and we do not include them in the fit procedure.

The combined expected negative log-Likelihood scan is shown in Fig.~\ref{fig:comb_kl}. The expected precision for the single channels is also shown. For completeness, we introduced in the combination also the \bbllll\ channel, which provides a sensitivity similar to the 4b channel. This decay channel was not re-optimized in this study and the result of the analysis is documented in Ref~\cite{L.Borgonovi:2642471}. The expected combined precision on the Higgs self-coupling obtained after combining the channels \bbaa, \bbtata, \bbbb\ and \bbllll\ can be inferred from the intersection of black curves with the horizontal 68\% and 95\% CL lines. The expected statistical precision for Scenario I, neglecting systematic uncertainties,  can be read from the dashed black line in Fig.~\ref{fig:comb_kl}, and gives $\dkl = 3.0$\% at 68\% CL. The  solid line corresponds to scenario II, while the boundaries of the shaded area represent respectively the alternative scenarios I and III. From the shaded black curve one can infer the final precision when including systematic uncertainties. Depending on the assumptions, the expected precision for the Higgs self-coupling is $\dkl = 3.4-7.8$\% at 68\% CL. The signal strength and self-coupling precision for the combination are summarized in Table~\ref{tab:comb_res}. 

The expected precision on the Higgs self-coupling as a function of the integrated luminosity is shown in Fig.~\ref{fig:comb_lumi}, for the three scenarios, with and without systematic uncertainties. With the most aggressive scenario I, a precision of $\dkl=10\%$ can be reached with only  3\,\iab{} of integrated luminosity, whereas approximately 20~\iab{} are required for the most conservative scenario III. Therefore, assuming scenario I, the 10\% target should therefore be achievable during the first 5 years of FCC-hh operations, combining the datasets of two experiments. Even including the duration of the FCC-ee phase of the project, and the transition period from FCC-ee to FCC-hh, this timescale is competitive with the time required by the proposed future linear colliders, which to achieve this precision need to complete their full programme at the highest beam energies.

\begin{figure}[ht!]
\centering     
  \subfigure
  {
   \includegraphics[width = 0.80\linewidth]{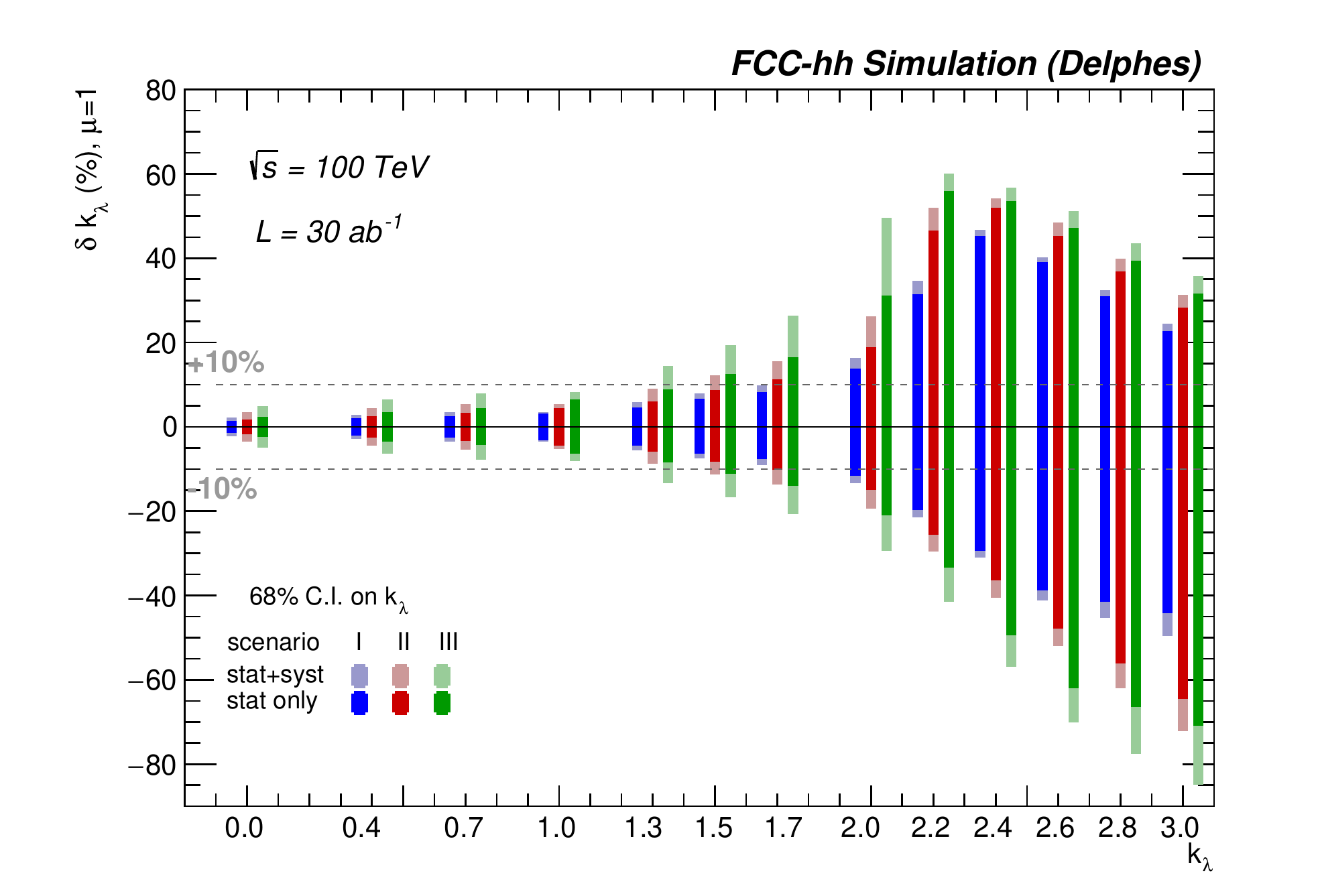}
  }
  \caption{Expected precision on the Higgs self-coupling as a function of the \kl value for each scenario. To improve readability, the positions of scenario I (III) bands are slightly offset in the negative (positive) direction along the \kl axis.}
  \label{fig:precisionkl}
\end{figure}

As already discussed, the value of the self-coupling coupling can significantly alter both the Higgs pair production cross section and the event kinematic properties. In order to explore the sensitivity to possible BSM effects in Higgs pair production, a multivariate BDT discriminant was optimised against the backgrounds for several values of \kl in the range $0<\kl<3$, in order to maximise the achievable precision for values of $\kl\neq1$. The BDT training has been performed only for the \bbaa\ channel, which dominates the overall sensitivity, whereas for the other channels we conservatively employ the BDT trained at $\kl=1$. The obtained precision as a function of $\kl$ is shown in Fig.~\ref{fig:precisionkl}~\footnote{We stress once more that, as discussed in Section~\ref{sec:theory}, precision projections for $\kl\ne 1$ are tied to a scenario in which only $\lthree$ is modified, and other BSM effects on the HH cross section are assumed to be negligible. For a recent study of the BSM modifications to kinematical distributions in presence of multiple anomalous couplings, see Ref.~\cite{Capozi:2019xsi}.}. 

 It can be seen that the overall precision follows the behaviour of the HH production cross section as function of \kl given in Figure~\ref{fig:xsec_vs_lambda}.  The best precision, \dkl$\approx$~2\%, is reached at $\kl=0$ where the value \slopebsm\ is large.  Conversely, the maximum uncertainty \dkl$\approx$~60\% is obtained at $\kl\approx2.5$,  and corresponds to the minimum of the total HH cross section, where \slopebsm\ $\to 0$ . As can be expected, the likelihood function presents a broad second minimum\footnote{The first minimum being at the probed value of \kl} in correspondence of the minimum of the HH cross section at $\kl=2.5$. The presence of this minimum is the reason behind the asymmetric behaviour of the uncertainties for the points near $\kl=2.5$. If the measurement is performed close enough to $\kl=2.5$ the likelihood falls in the second minimum before reaching the 68\% C.L. threshold, thus enlarging the measurement uncertainty in one direction. It should be noted that, while the HH cross section is roughly symmetric around $\kl=2.5$, we do not expected the uncertainties to be symmetric as well, as the kinematic behaviour of the HH system are quite different between $\kl<2.5$ and $\kl>2.5$. It can also be noticed that when switching on the systematic uncertainties the precision at small \kl degrades compared to the SM case. This reflects the fact that the HH kinematics at $\kl\approx0$ are similar to the single-Higgs background.

\begin{table}
\centering
\begin{tabular}{cccc}

 @68\% CL   & scenario I & scenario II & scenario III \\
\hline 
\dmu\,\,\,\,  
\begin{tabular}{ll}     stat only  \\    stat + syst \end{tabular}
&
    \begin{tabular}{@{}c@{}} 2.2 \\ 2.4 \end{tabular}
&
    \begin{tabular}{@{}c@{}} 2.8 \\ 3.5 \end{tabular}
&
    \begin{tabular}{@{}c@{}} 3.7 \\ 5.1 \end{tabular}
\\
\hline 

\dkl\,\,\,\, 
\begin{tabular}{ll}     stat only  \\    stat + syst \end{tabular}
&
    \begin{tabular}{@{}c@{}} 3.0 \\ 3.4 \end{tabular}
&
    \begin{tabular}{@{}c@{}} 4.1 \\ 5.1 \end{tabular}
&
    \begin{tabular}{@{}c@{}} 5.6 \\ 7.8 \end{tabular}

\end{tabular}
\caption{\label{tab:comb_res} Combined expected precision at 68\% CL on the di-Higgs production cross- and Higgs self coupling using all channels at the FCC-hh with \intlumifcc. The symmetrized value $\delta=(\delta^++\delta^-)/2$ is given in \%.}
\end{table}

\section{Conclusions and perspectives}
\label{sec:conc}
The precise measurement of the Higgs self-coupling must be a top priority of future high-energy collider experiments. 
Previous studies on the potential of a 100~TeV pp collider have discussed the sensitivity of various decay channels, often based on simple rectangular cut-based analyses~\footnote{Just before the public release of this work, we learned of a similar study presented in Ref.~\cite{Park:2020yps}, using a multivariate analysis of the \bbaa\ final state. While many aspects of the two studies are different, in particular for what concerns the consideration of systematic uncertainties, there is quantitative agreement on the improvements induced by the use of multivariate analysis.}. 
In the present study the measurement strategy has been optimized in the \bbaa, \bbtata\ and \bbbb\ channels using machine learning techniques. For the first time, a precise set of assumptions of detector performances and possible sources of systematic uncertainties has been defined and used to derive the achievable precision. Consistently with our previous findings, the \bbaa\ channel drives the final sensitivity, with an expected precision of $\dkl= 3.8-10.0\%$ depending on the detector and systematic assumptions. The \bbtata\ and \bbbb\ channels provide instead a less precise single channel measurement, respectively of $\dkl= 10-14\%$ and $\dkl= 22-32\%$.  

The final combined sensitivity across all considered channels leads to an expected precision at the FCC-hh on the Higgs self-coupling $\dkl= 3.4-7.8\%$ with an integrated luminosity of \intlumifcc.
By considering the most agressive detector and systematics assumptions, the 10\% threshold can be achieved with $\sim 3$\,\iab, corresponding to $\sim 3-5$ years of early running at the start-up luminosity of $5\times 10^{34}$\,\ilum.

This work shifts the perspective on the Higgs self-coupling ultimate precision at FCC from being  statistics dominated to systematics dominated. This is a crucial new development: on one side it gives us confidence that the design parameters of FCC-hh are well tailored to reach, unique among all proposed future collider facilities, the few-percent level of statistical precision. On the other hand, it calls for a more thorough assessment of all systematic uncertainties. For example, we should validate the estimates presented in this work through full simulations of more realistic detector designs in the presence of pile-up, and explore all other possible handles to further reduce them. The huge FCC statistics will provide multiple control samples, well beyond what discussed in our paper, that could be used for these purposes, and in particular to pin down the background rates with limited reliance on theoretical calculations.
At this level of precision, however, the theoretical uncertainties on the HH signal will play an important role in the extraction of the self-coupling from the measured production rate. As indicated in Section~\ref{subsec:systematics}, we assumed in our study an uncertainty on the HH cross sections ranging from 0.5 to 1.5\%. This would need the theoretical predictions to improve relative to today's knowledge, requiring the extension of the perturbative order by {\it at least} one order beyond today's known NLO with full top mass dependence~\cite{Borowka:2016ypz}, and possibly beyond the N$^3$LO in the $m_{top}\to \infty$ limit, recently achieved~\cite{Chen:2019lzz,Chen:2019fhs}.
This will be very challenging, and it is impossible today to estimate the asymptotic reach in theoretical precision. Nevertheless, the innovative technical progress we have witnessed in the recent years encourages us to assume that the necessary improvements are possible within the several decades that separate us from the first FCC-hh run. 

In conclusion, this study strengthens the evidence that a 100~TeV pp collider, with integrated luminosity above 3\,\iab, can measure the Higgs self-coupling more precisely than any other proposed project, on a competitive time scale.


\begin{acknowledgments}

We would like to thank the FCC group at CERN. In particular we thank Alain Blondel and Patrick Janot for helpful suggestions, Clement Helsens, Valentin Volkl and Gerardo Ganis for the valuable help and support on the FCC software. We also thank Gudrun Heinrich and Stephen Jones for helpful discussions on NLO Higgs pair event generation. Finally, we would like to acknowledge the \bbllll\ channel study from Lisa Borgonovi, Elisa Fontanesi  and Sylvie Brabant, which we have used as one of the inputs for our determination of the self-coupling combined sensitivity.
\end{acknowledgments}




\bibliographystyle{JHEP}
\clearpage
\bibliography{paper}
\clearpage

\appendix

\section{Statistical procedure}
\label{subsec:procedure}
The statistical methodology used in this paper relies on the strategy adopted by the ATLAS and CMS Collaborations, and described in Ref.~\cite{LHC-HCG}.
A detailed description of the procedures used in this paper are described in more detail in Refs.~\cite{Chatrchyan:2012tx, Chatrchyan:2012ufa}.
The {\sc Combine} software package~\cite{combine} has been used as statistical and fitting tool to produce the final results. {\sc Combine} is based on the standard LHC data modeling and handling toolkits {\sc RooFit}~\cite{Verkerke:2003ir} and {\sc RooStats}~\cite{Moneta:2010pm} and it is developed and maintained by the CMS collaboration.

The parameter of interest (POI) tested in these results are either the trilinear coupling modifier \klfrac\ or the double Higgs signal strength \mufrac, defined as the ratio between the (expected) measured double Higgs yield and its SM expectation.

In the model, the POI $\alpha=k_\lambda$ or $\alpha=\mu$ is estimated with its corresponding confidence intervals using a profile likelihood ratio test statistic $q(\alpha)$~\cite{Cowan:2010st,LHC-HCG}, in which experimental or theoretical uncertainties are incorporated via nuisance parameters (NP). Given a of POI $\alpha$ that depends on the set of NP $\vec\theta$, $q(\alpha)$ is defined as:

\begin{equation}
  q(\alpha) = -2\ln \left(\frac{L\big(\alpha\,,\,{\hat{\vec\theta}}_{\alpha}\big)}
{L(\hat{\alpha},\hat{\vec\theta})\label{eq:LH}}\right).
\end{equation} 
An individual NP represents a single source of systematic uncertainty. Its effect is therefore considered fully correlated between all of the the final states included in the fit that share a dependency on it, as will be discussed later in this section.

The quantities $\hat{\alpha}$ and $\hat{\vec\theta}$ denote the unconditional maximum likelihood estimates of the parameter value, while $\hat{\vec\theta}_{\alpha}$ denotes the conditional maximum likelihood estimate for fixed values of the POI $\alpha$.

The likelihood functions in the numerator and denominator of Eq.~(\ref{eq:LH}) are constructed using products of probability density functions (PDFs) of signal and background for the various discriminating variables used in the input analyses, as well as constraint terms for the NPs. The PDFs are built from the BDT distributions described in Section \ref{sec:analysis}. It should be noted that while the signal shape depends on the value on \kl, that dependence is relatively mild. Given the expected precision of $\mathcal{O}(10\%)$ on the measurement of $k_\lambda$ at the FCC-hh, the effect of its variation on the signal lineshape is minimal when performing the measurement at a given value of \kl and can be safely neglected. The effects on acceptance and selection efficiencies of varying $k_\lambda$ or $\mu$ are instead considered in the fit. The expected precision on \kl\ and $\mu$ is assessed by performing a likelihood fit on a a pseudo-data set that has been constructed assuming $\mu=1$ and $k_\lambda=1$, using the asymptotic approximation as described in~\cite{Cowan:2010st,Junk:2016jag}.





\clearpage

\end{document}